\documentclass[sigconf, nonacm]{acmart}

\newcommand\vldbdoi{XX.XX/XXX.XX}

\newcommand\vldbvolume{17}
\newcommand\vldbissue{3}

\newcommand\vldbavailabilityurl{URL_TO_YOUR_ARTIFACTS}

\newcommand\vldbpagestyle{plain} 

\usepackage{graphicx}   
\usepackage{algorithm}
\usepackage{multirow}

\usepackage{graphicx}
\usepackage{subcaption}

\usepackage{booktabs}
\usepackage{paralist}
\usepackage{balance}
\usepackage{siunitx}
\usepackage[normalem]{ulem}
\useunder{\uline}{\ul}{}

\usepackage[noend]{algpseudocode}
\usepackage{amsmath}
\usepackage{setspace}
\usepackage[justification=centering]{caption}
\usepackage{subcaption} 
\usepackage{xcolor}
\usepackage{mathtools}
\usepackage{appendix}

\usepackage[shortlabels]{enumitem}
\setenumerate[1]{itemsep=0pt,partopsep=0pt,parsep=\parskip,topsep=0.5pt}
\setitemize[1]{itemsep=0pt,partopsep=0pt,parsep=\parskip,topsep=0.5pt}
\setdescription{itemsep=0pt,partopsep=0pt,parsep=\parskip,topsep=0.5pt}

\newcommand\MPIN{MPIN}

\newcommand\PM{MPIN-P}
\newcommand\DU{MPIN-D}
\newcommand\MU{MPIN-M}
\newcommand\DMU{MPIN-DM}

\newcommand\WIFI{Wi-Fi}
\newcommand\ICU{ICU}
\newcommand\AIR{Airquality}

\newcommand{\best}[1]{\textbf{#1}}

\newcommand{\secBest}[1]{\uline{#1}}

\newcommand\ExpCaption[1]{%
     \captionsetup{font=small}%
     \caption{#1}}

\newtheorem{lemma}{\bf Lemma}
\newtheorem{definition}{\bf Definition}

\newcommand{\Xiao}[1]{{\textcolor{red} {#1}}}

\newtheorem*{theorem*}{\bf Research Problem}
\newtheorem{example}{\bf Example}

\newtheorem{criterion}{Criterion} 

\usepackage[colorinlistoftodos,prependcaption,textsize=tiny]{todonotes}

\begin{document}

\title{Missing Value Imputation for Multi-attribute Sensor Data Streams via Message Propagation (Extended Version)}


\author{Xiao Li$^1$ \ \ \ \ \ \ Huan Li$^2$ \ \ \ \ \ \ Hua Lu$^1$\ \ \ \ \ \ Christian S. Jensen$^3$\ \ \ \ \ \ Varun Pandey$^4$\ \ \ \ \ \ Volker Markl$^{4,5,6}$}
\affiliation{%
  \institution{$^1$Department of People and Technology, Roskilde University, Denmark}
  \institution{$^2$College of Computer Science and Technology, Zhejiang University, China}
  \institution{$^3$Department of Computer Science, Aalborg University, Denmark}
  \institution{$^4$Technische Universität Berlin, Germany \hspace{0.1cm}  $^5$DFKI Berlin, Germany \hspace{0.2cm}  $^6$BIFOLD, Germany}
  \institution{$^1$\{xiaol,  luhua\}@ruc.dk \hspace{0.2cm} $^2$lihuan.cs@zju.edu.cn \hspace{0.2cm} $^3$csj@cs.aau.dk \hspace{0.1cm}  $^4$\{varun.pandey,  volker.markl\}@tu-berlin.de}
}

\renewcommand{\authors}{Xiao Li, Huan Li, Hua Lu, Christian S. Jensen, Varun Pandey, and Volker Markl}

\begin{abstract}
Sensor data streams occur widely in various real-time applications in the context of the Internet of Things (IoT). However, sensor data streams feature missing values due to factors such as sensor failures, communication errors, or depleted batteries. Missing values can compromise the quality of real-time analytics tasks and downstream applications. Existing imputation methods either make strong assumptions about streams or have low efficiency. 
In this study, we aim to accurately and efficiently impute missing values in data streams that satisfy only general characteristics in order to benefit real-time applications more widely. First, we propose a message propagation imputation network (MPIN) that is able to recover the missing values of data instances in a time window. We give a theoretical analysis of why MPIN is effective. Second, we present a continuous imputation framework that consists of data update and model update mechanisms to enable MPIN to perform continuous imputation both effectively and efficiently. Extensive experiments on multiple real datasets show that MPIN can outperform the existing data imputers by wide margins and that the continuous imputation framework is efficient and accurate.

\end{abstract}

\pagestyle{plain} 
\settopmatter{printfolios=true}

\maketitle

\pagestyle{\vldbpagestyle}

\begingroup
\renewcommand\thefootnote{}\footnote{\noindent
This work is licensed under the Creative Commons BY-NC-ND 4.0 International License. Visit \url{https://creativecommons.org/licenses/by-nc-nd/4.0/} to view a copy of this license. For any use beyond those covered by this license, obtain permission by emailing \href{mailto:info@vldb.org}{info@vldb.org}. Copyright is held by the owner/author(s). Publication rights licensed to the VLDB Endowment. \\
\raggedright Proceedings of the VLDB Endowment, Vol. \vldbvolume, No. \vldbissue\ %
ISSN 2150-8097. \\
\href{https://doi.org/\vldbdoi}{doi:\vldbdoi} \\
}\addtocounter{footnote}{-1}\endgroup


\section{Introduction}

With the increasing deployment of the Internet of Things (IoT), \emph{multi-attribute} sensor data streams (also known as {multi-attribute} time series) can be found in numerous application domains, including in medical services~\cite{che2018recurrent,yoon2019nets}, meteorology~\cite{tran2012claro,chen2002multi}, transportation~\cite{abadi2003aurora,mirylenka2015conditional}, and energy~\cite{sharaf2004balancing}.
For example, in the Intensive Care Unit (ICU) of a hospital, medical professionals may need to continuously monitor patients' health status through a system that tracks vital signs such as heart rate, blood pressure, and body temperature. As another example, it is imperative for an air quality monitoring system to reliably track diverse metrics such as PM2.5 and SO2, across multiple locations in a city. Such systems produce data streams, i.e., continuous and unbounded sequences of data instances. Each data instance in turn is characterized by a vector of attribute values, as illustrated in Figure 1. In the context of the ICU scenario, a data instance is a vector of health-related values such as heart rate, blood pressure, and body temperature. These values typically capture information about the condition of a patient using multiple sensors at a given point in time.

In real-world systems, however, data streams may contain missing values in their instances due to factors such as sensor failures, depleted batteries, and communication errors~\cite{arous2019recovdb,ren2019skyline,whang2020data}. All such factors can result in data instances with missing values (see observed and missing attributes in Figure 1).
%
%
\begin{figure}[!ht]
\centering
\includegraphics[width=0.9\columnwidth]{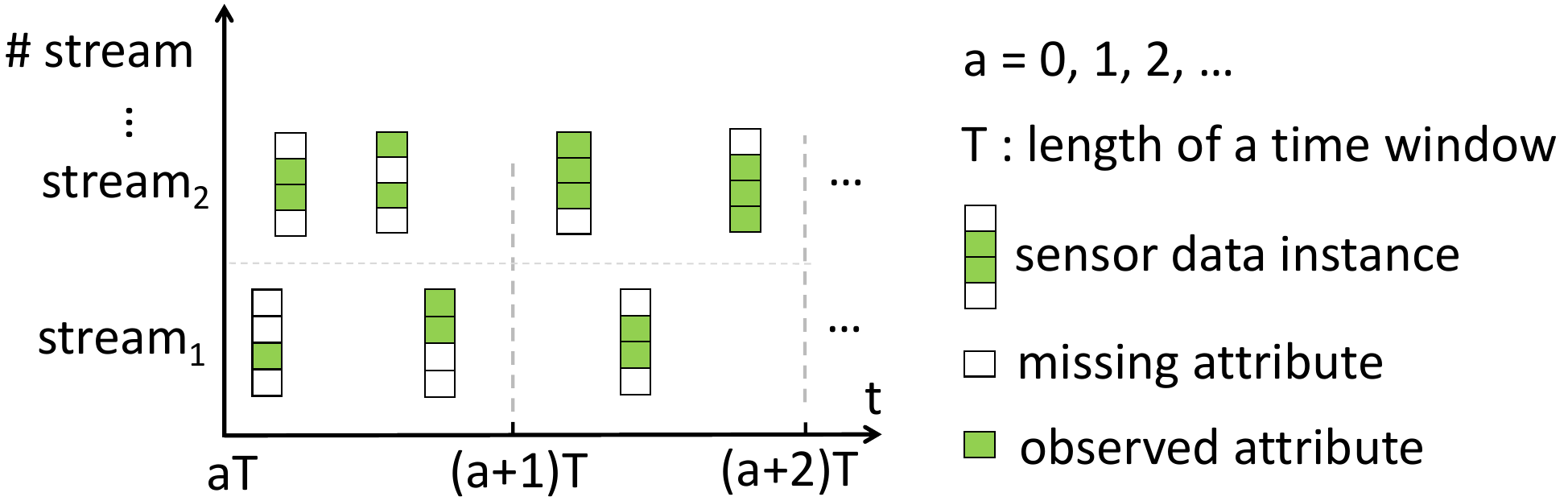}
\caption{Sensor data streams.}
\label{fig:sen_streams}
\end{figure}
Despite the fact that a system might continue to function with partially observed data, inaccurate or ill-defined results~\cite{berti2018discovery,cambronero2017query,ren2019skyline} may be produced. 
To make it worse, the sparsity can be quite high. For example, in  ICU and Wi-Fi datasets that we use in this paper, the data sparsity exceeds 80\% (see Table~\ref{tab:dataset_des}).
The presence of high sparsity in data streams can cause major issues for online analytical tasks and downstream applications. In the ICU example, a doctor may make an incorrect, life-critical conclusion about the health of a patient based on monitored vital signs with missing values. 
In the example of air quality monitoring, missing values in the data streams may cause the system to miss the critical conditions of a fire emergency and consequently fail to give timely alerts.
%
%
Therefore, it is crucial to accurately recover missing values in data streams in real time.

However, it is nontrivial to do so given that sensor data streams may exhibit characteristics as follows.
\textbf{C1 (Aperiodicity)}: Data instances occur at non-fixed intervals.
\textbf{C2 (Concurrency)}: Concurrent streams exist, such as multiple air quality monitoring stations or simultaneous monitoring of multiple patients in an ICU scenario.
\textbf{C3 (Heterogeneity)}: Data instances contain diverse attributes, such as body temperature and blood pressure in an ICU scenario.
\textbf{C4 (Sparsity)}: Streams often exhibit high ratios of missing values.

Existing studies on data stream imputation either have low efficiency~\cite{cao2018brits,du2023saits,andrea2021filling} or make strong assumptions on the homogeneity of streams~\cite{papadimitriou2005streaming,yu2016temporal,mei2017nonnegative}. 
Such methods generally fall into two categories.
First, neural network-based imputers~\cite{cao2018brits,du2023saits,andrea2021filling} employ sequential neural networks, e.g., Recurrent Neural Networks (RNN), to iteratively fill in missing values within a sequence in a transductive learning mechanism. Such a sequential structure-based model often cannot contend with characteristics C1 and C2 --- the aperiodicity of streams can prevent a model from capturing the temporal dependencies for data imputation, while the concurrency of streams can render the imputation inefficient as the model needs to be trained sequentially on all streams.

Second, matrix-based imputers~\cite{papadimitriou2005streaming,yu2016temporal,mei2017nonnegative} focus solely on homogeneous, single-attribute streams and they mostly fail to contend with characteristics C3 and C4. The heterogeneity of attributes and high ratios of missing values may render matrix-based imputers less effective, as values in a matrix may carry different meanings, and high sparsity in a matrix can compromise matrix completion operations that attempt to recover missing values.

To address these challenges, we propose a set of techniques to overcome the problems in a tumbling window. First of all, we construct a similarity graph to link data instances that originate from different streams or different timestamps within a time window. 
We do not assume that a stream is periodic; neither do we need to know from which stream an instance originates. 
We primarily construct a graph based on the similarities of data instances, as we believe this to be most significant for imputation. 
Also, compared to matrix-based imputers~\cite{papadimitriou2005streaming,yu2016temporal,mei2017nonnegative}, operating on a graph can benefit from positive relational biases~\cite{battaglia2018relational}. Namely, an instance only needs to be related to the most similar instances instead of to all other instances. Such a positive bias has been demonstrated to be beneficial to missing value imputation~\cite{cini2021multivariate}. 

Subsequently, we propose a message propagation process on the similarity graph and design a Message Propagation Imputation Network (\MPIN{}) that can exploit the correlations among instances and attributes to impute missing values accurately. We also provide a theoretical study for why MPIN is more effective in exploiting correlations for data imputation than a recently proposed feature propagation process~\cite{rossi2022unreasonable}. Although \MPIN{} conducts imputation in a  transductive learning fashion, as do existing neural network-based imputers, \MPIN{} is significantly more time-efficient since it is a graph-based model and can conduct imputation on multiple graph nodes (i.e., instances) in parallel, thus being able to exploit available computing resources sufficiently.  

 Although \MPIN{} is  both effective and efficient at imputing missing values of data instances in a time window, there are other challenges when \MPIN{} is applied to continuous imputation for data streams. Essentially, the aperiodicity of data streams causes the number of data instances to vary from window to window. Consequently,  there may not be enough data for training \MPIN{} effectively. To contend with this, we propose a data update mechanism that can keep and update the important data instances in the streams and utilize them to enable more effective continuous imputation. 
Furthermore, as \MPIN{} relies on transductive learning, retraining is needed at every current time window in order to impute newly arriving data instances.  In order to lower the retraining cost, we propose a model update mechanism that allows to resume training from the best model state along the timeline so far. This makes continuous imputation with \MPIN{} more efficient.

In summary, we make the following contributions.
\begin{itemize}[leftmargin=*]
    \item We construct a similarity graph with data instances in a time window having data instances as graph nodes. In addition, we propose a message propagation process on the graph to enable capturing correlations and exploiting positive relational bias. 
    
    \item Based on the message propagation process, we propose a message propagation imputation network (\MPIN{}) to exploit correlations among instances to impute their missing values in a time window. We also give a theoretical analysis of why \MPIN{} is effective.
    \item  To use \MPIN{} for continuous imputation, we propose a framework with data update and model update mechanisms that help achieve both effective and efficient continuous imputation.

    \item We report on extensive experiments showing that the proposed \MPIN{} can outperform competitors at data imputation and that the continuous imputation framework is effective and efficient.
\end{itemize}

The rest of the paper is organized as follows.
Section~\ref{sec:preliminary} presents preliminaries and the research problem. Section~\ref{sec:proposed_model} details the \MPIN{} model for snapshot data imputation in a time window. Section~\ref{sec:incremental} presents an \MPIN{} based continuous imputation framework. Section~\ref{sec:experiments} reports on extensive experiments. Section~\ref{sec:related} reviews related work.  Section~\ref{sec:conclu} concludes and covers future research.
\vspace*{-5pt}

\section{Preliminaries and Problem }\label{sec:preliminary}
Table~\ref{tab:notation} presents commonly used notation.

\begin{table}[!htbp]
\centering
\small
\caption{Notation.}
\label{tab:notation}
\begin{tabular}{c|l}
\toprule
{Symbol} & {Description} \\ \midrule
$\mathbf{x}$ & sensor data instance \\
$\mathbf{m}$ & mask of a sensor data instance \\
$\mathcal{X}$ & sensor data streams \\
$\mathcal{M}$ & mask of sensor data streams \\
$\mathcal{X}_a$ & sensor data chunk of a time window \\
$\mathcal{M}_a$ &mask of a sensor data chunk  \\
\bottomrule
\end{tabular}
\vspace*{-5pt}
\end{table}

\subsection{Data and Notation}

\begin{definition}[Sensor Data Instance]
A \emph{sensor data instance} is represented as a $\mathtt{D}$-dimensional vector $\mathbf{x} \in \mathbb{R}^\mathtt{D}$, where each dimension (i.e., attribute) $\mathbf{x}[d]$ ($0 \leq d < D$) captures a sensor measurement.
\end{definition}

When clear from the context, we use  "instance" to refer to a sensor data instance.
An instance may consist of homogeneous or heterogeneous sensor measurements.
In the aforementioned ICU example, an instance refers to a vector of heterogeneous health-related measures such as heart rate and blood pressure.

\begin{definition}[Sensor Data Streams]
A \emph{sensor data stream} is an unbounded,  time-ordered sequence of sensor data instances. $\mathtt{J}$ concurrent sensor data streams are organized as a tensor $\mathcal{X}$ of size $\mathtt{J} \times  \mathtt{N} \times \mathtt{D} $ such that $\mathtt{N} \rightarrow \infty$ corresponds to the time dimension.
We use $\mathcal{X}^{j} = \mathcal{X}[j,:,:]$ to denote the $j$-th ($0 \leq j < \mathtt{J}$) data stream and $\mathcal{X}^{j}(n) = \mathcal{X}[j,n,:]$ to denote the $n$-th ($0 \leq n < \mathtt{N}$) sensor data instance of the $j$-th data stream.\footnote{In general, $n$ can be a timestamp or an index (sequence number) but it takes only one form in a given data stream. Meanwhile, a stream is either periodic or aperiodic.}
\end{definition}

The instances in a stream may contain \emph{missing values} in their attributes due to factors such as sensor failures, depleted batteries, communication errors, and unforeseen malfunctions~\cite{arous2019recovdb,ren2019skyline,whang2020data}.
To indicate the missing values in sensor data streams, we define the notion of mask for sensor data streams as follows.

\begin{definition}[Mask for Sensor Data Streams]
Given the sensor data streams $\mathcal{X}$, its corresponding \emph{mask} $\mathcal{M}$ is a binary tensor with the same shape as $\mathcal{X}$:
\begin{equation}
\mathcal{M}[j,n,d] = 
\begin{cases}
0, \;\;\;\mathcal{X}[j,n,d]~\text{is missing} ; \\
1, \;\;\;\text{otherwise}.
\end{cases}
\end{equation}
\end{definition}

Typically, a mask $\mathcal{M}$ is sparse with many zeros.
The sparsity characteristic of sensor data streams often renders online analytical tasks or downstream real-time applications (e.g., air quality monitoring) inaccurate or even non-functional~\cite{berti2018discovery,cambronero2017query,ren2019skyline}.
%

\vspace*{-5pt}

\subsection{Research Problem Formulation}
\label{ssec:problem_formulation}

Data streams are often processed using \textbf{tumbling window} technique~\cite{patroumpas2006window, traub2019efficient}.
Specifically, a tumbling window is a fixed-length time window that moves through a stream at a constant time interval $\mathtt{T}$ in a non-overlapping fashion.
With such windows, we discretize the sensor data streams $\mathcal{X}$ into data chunks and represent each chunk as $\mathcal{X}_a = \mathcal{X}[:, a\mathtt{T}:(a+1)\mathtt{T}, :]$, where $a \in \mathbb{Z}^{0+}$. 
For ease of presentation, we assume that 
each time window contains $\mathtt{T}$ time units and at most $\mathtt{T}$ data instances. Formally, $\mathcal{X}_a \in \mathbb{R}^{\mathtt{J} \times \mathtt{T} \times \mathtt{D}}$.  
%

To further facilitate actual processing, we convert a data chunk $\mathcal{X}_a$ into a 2-dimensional matrix $\mathcal{X}_a \in \mathbb{R}^{(\mathtt{J} \cdot \mathtt{T}) \times \mathtt{D}} $ by merging streams (i.e., $\mathtt{J}$) and time (i.e., $\mathtt{T}$). As a result, a \textbf{data chunk} $\mathcal{X}_a$ encompasses at most $\mathtt{J} \cdot \mathtt{T}$ data instances, to which each of the  $\mathtt{J}$ streams contribute at most $\mathtt{T}$  instances in a time window.
The corresponding mask can be discretized likewise, resulting in a mask chunk $\mathcal{M}_a$ ($a \in \mathbb{Z}^{0+}$) for the corresponding time window $[a \mathtt{T}:(a+1)\mathtt{T}]$.

We proceed to define the research problem.
\begin{theorem*}[Imputation of Sensor Data Streams, IDS]
Given sensor data streams $\mathcal{X}$, \emph{online imputation of sensor data streams} continuously takes the current data chunk $\mathcal{X}_a$ as input and returns a complete data chunk $\hat{\mathcal{X}}_a$ by replacing each missing value in $\mathcal{X}_a$ with a proper data value.
The objective here is to recover those missing values in $\mathcal{X}_a$ accurately and efficiently.
\end{theorem*}


We solve the IDS problem at two different, yet correlated, levels. First, we focus on snapshot imputation for a single time window, i.e., imputing missing values of data instances in one data chunk. In Section~\ref{sec:proposed_model}, we construct a similarity graph to capture the correlations among data instances and propose a Message Propagation Imputation Network (MPIN) that exploits the graph to impute the missing values in the current data chunk. Second, we study efficient and effective imputation for continuous time windows. In Section~\ref{sec:incremental}, we design a framework that uses MPIN as a building block for continuous imputation for data chunks from consecutive windows.

\section{Snapshot Imputation for a Window}
\label{sec:proposed_model}

This section focuses on snapshot imputation for a single time window. The key innovation is the Message Propagation (\textsc{MsgProp}) Imputation Network (MPIN) that takes a data chunk with missing values as input and outputs a data chunk without missing values. MPIN works on a similarity graph that captures the correlations among data instances in the window. MPIN's \textsc{MsgProp} component extends and generalizes the recent feature propagation~\cite{rossi2022unreasonable} (\textsc{FeaProp}) for graph node feature imputation.
Section~\ref{ssec:similarity_graph} presents the similarity graph, 
Section~\ref{ssec:fea_vs_msg} compares \textsc{MsgProp} and \textsc{FeaProp}, and Section~\ref{ssec:imputation_network} details \MPIN{}.

\subsection{Similarity Graph}
\label{ssec:similarity_graph}

\begin{figure}[!ht]
\centering
\includegraphics[width=0.85\columnwidth]{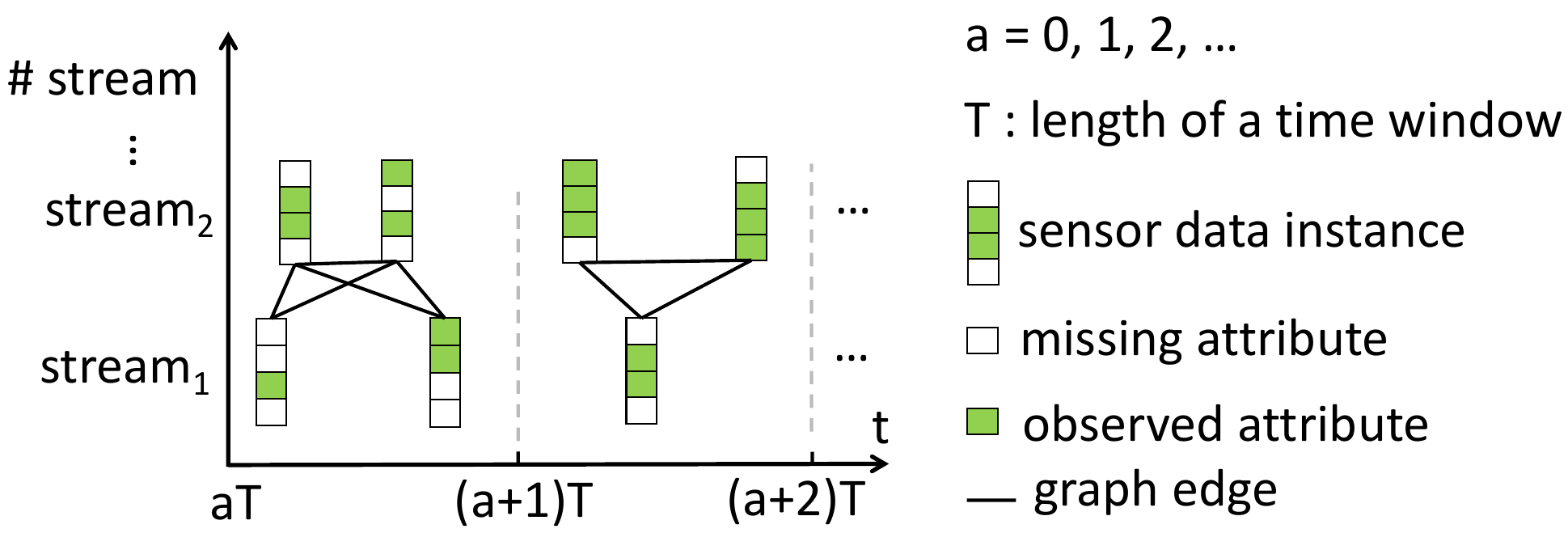}
\caption{Similarity graphs for different data chunks.}
\label{fig:simi_graph_wind}
\vspace*{-5pt}
\end{figure}

Given a data chunk $\mathcal{X}_a  \in \mathbb{R}^{(\mathtt{J} \cdot \mathtt{T}) \times \mathtt{D}}$, we construct a \emph{similarity graph} $G=(V, E)$ to organize $\mathcal{X}_a$'s data instances. Figure~\ref{fig:simi_graph_wind} illustrates similarity graphs.
Specifically, each data instance $\mathbf{x}_i$ ($0 \leq i < \mathtt{J} \cdot \mathtt{T}$) corresponds to a graph node in $V$, where  $|V| = \mathtt{J} \cdot \mathtt{T}$.
For each pair of instances $\mathbf{x}_i$ and $\mathbf{x}_k$, we determine whether they should be linked by an edge $e_{ik}$ as follows.
First, we fill-in each missing value of each instance using the mean of the observed values in the same dimension of all other instances in $V$.
Second, we compute the similarity between $\mathbf{x}_i$ and $\mathbf{x}_k$ and then create an edge between them only if they are sufficiently similar.
The similarity can be implemented based on Euclidean distance or Cosine distance~\cite{wang2002clustering}. For example, we may simply check whether the Euclidean or Cosine distance between the vectors of $\mathbf{x}_i$ and $\mathbf{x}_k$ is below a given threshold. Alternatively, we may check whether $\mathbf{x}_i$ is among $\mathbf{x}_k$'s $K$ nearest neighbors (KNN) in terms of  Euclidean or Cosine distance, or vice versa. 
{As our preliminary experiments in Section~A.3 in Appendix show, similarity graphs built using the Euclidean distance-based KNN method achieve the best performance. Therefore, we adopt this method in the experiments.}

The similarity graph correlates data instances from different timestamps (but within the same time window) or across different streams. This offers a view of relationships among data instances, that is broader than if we focus on a single stream only. 
The graph also allows us to design a message propagation mechanism to exploit correlations among nodes (i.e., instances) and impute the missing attribute values for them.

\vspace*{-5pt}
\subsection{\textsc{FeaProp} versus \textsc{MsgProp}}\label{ssec:fea_vs_msg}

Given the similarity graph $G(V,E)$, we obtain a $|V| \times |V|$ adjacency matrix $\mathbf{A}$.
Formally, $\mathbf{A}[i,k] = 1$ if graph edge $e_{ik}$ exists and $0$, otherwise.
Subsequently, we compute the diagonal degree matrix $\mathbf{D} = \operatorname{diag}(\sum\nolimits_i \mathbf{A}[i,1], \ldots, \sum\nolimits_i \mathbf{A}[i,|V|])$ and the normalized adjacency matrix $\Tilde{\mathbf{A}} = \mathbf{D}^{-\frac{1}{2}}\mathbf{A}\mathbf{D}^{-\frac{1}{2}}$.

\noindent\textbf{Feature Propagation}. {Given graph $G$,  the \textsc{FeaProp} process aims to minimize the Dirichlet energy of the graph, i.e., $\min \sum_{ik} \big( \Tilde{\mathbf{A}}[i,k](\mathbf{x}_i - \mathbf{x}_k)^2 \big)$. This is a widely used approach to promote feature homophily in graphs~\cite{rossi2022unreasonable}. Essentially, it aims to make a node's feature similar to those of its neighboring nodes. For instance, if $\Tilde{\mathbf{A}}[i,k]=1$, $\mathbf{x}_i$ and $\mathbf{x}_k$ are adjacent nodes. As a result, $x_i$ and $x_k$ should be similar in order to make $(\mathbf{x}_i - \mathbf{x}_k)^2$ small, thus minimizing the objective function. This way, the missing values of $\mathbf{x}_i$ can be interpolated from the corresponding values of $\mathbf{x}_k$ or vice versa. }
As the Dirichlet energy is convex, an efficient solution is the iterative weighted-sum of neighbors' information as follows.
\vspace*{-3pt}
\begin{equation}\label{equ:fp_solution}
\mathbf{\tilde{x}}_i^{(l+1)} = \sum\nolimits_{\mathbf{x}_k \in {\mathcal{N}_i}} \Tilde{\mathbf{A}}_{ik} \mathbf{x}_k^{(l)}, 
\end{equation}
\vspace*{-2pt}
where $\Tilde{\mathbf{A}}_{ik}$ refers to $\Tilde{\mathbf{A}}[i,k]$, $\mathbf{\tilde{x}}_i^{(l+1)}$ is the imputation result of $\mathbf{x}_i$ in the $(l+1)$-th iteration, and the set $\mathcal{N}_i$ contains all nodes that are adjacent to $\mathbf{x}_i$. 
Theoretical analyses can be found elsewhere~\cite{rossi2022unreasonable}. Originally, $\mathbf{\tilde{x}}_i^{(l+1)}$is directly taken into the right-hand side of Equation~\ref{equ:fp_solution} to start the next iteration, but this will modify the observed, correct values in $\mathbf{x}_i$ and cause a degradation in performance. 

To solve the potential problem, a bound condition on the reconstructed features is added, which keeps the values of observed features equal to their original values during iterations. The bound condition is computed as follows.
\begin{equation}\label{equ:fp_bound}
\mathbf{x}_i^{(l+1)} \gets \mathbf{x}_i^{(0)} \odot \mathbf{m}_i +  \mathbf{\tilde{x}}_i^{(l+1)} \odot (\textbf{1} - \mathbf{m}_i),
\end{equation}
where $\mathbf{m}_i$ is the corresponding part of the mask for $\mathbf{x}_i$ and $\odot$ is the element-wise product operator.

After a certain number of iterations according to Equations~\ref{equ:fp_solution} and~\ref{equ:fp_bound}, we obtain the converged feature $\hat{\mathbf{x}}_i$ for each data instance $\mathbf{x}_i$.
In \textsc{FeaProp}, $\hat{\mathbf{x}}_i$ is used as the reconstructed version of $\mathbf{x}_i$.
The proof of convergence of \textsc{FeaProp} is available elsewhere~\cite{rossi2022unreasonable}.

\smallskip
\noindent\textbf{Message Propagation}. 
{The \textsc{FeaProp} process utilizes iterative adjacency matrix multiplication with bound conditions to impute missing values of features of nodes (i.e., instances).  However,  the adjacency matrix (i.e., $\Tilde{A}$) treats a node's neighboring nodes equally, and this may not be accurate in reality as a node may have different correlations with different neighboring nodes. Next, the different attributes of an instance may also be correlated. Thus, we introduce two factors, namely correlations among instances (i.e., nodes) and among attributes of a node,  to enhance imputation performance.}
%
Specifically, we extend Equation~\ref{equ:fp_solution} in \textsc{FeaProp} as follows.
\begin{equation}\label{equ:mp_solution_1}
\mathbf{\tilde{x}}_i^{(l+1)} = \sum\nolimits_{\mathbf{x}_k \in {\mathcal{N}_i}} c(\mathbf{x}_i^{(l)}, \mathbf{x}_k^{(l)}) \mathbf{x}_k^{(l)} \mathbf{W}.
\end{equation}

Equation~\ref{equ:mp_solution_1} differs  from Equation~\ref{equ:fp_solution} in a number key points. First, Equation~\ref{equ:mp_solution_1} replaces the constant component of adjacency matrix $\Tilde{\mathbf{A}}_{ik}$ by a learnable correlation function $c(\mathbf{x}_i^{(l)}, \mathbf{x}_k^{(l)})$. Second, a feature transformation matrix $\mathbf{W} \in \mathbb{R}^{\mathtt{D} \times \mathtt{D}}$ is added to Equation~\ref{equ:mp_solution_1} in order to capture correlations among different attributes of $\mathbf{x}_i$.\footnote{We omit the bias vector for the sake of brevity.}
Above, the shape of $\mathbf{W}$ is restricted to $\mathtt{D} \times \mathtt{D}$ to make the output dimensions consistent with the input dimensions.
To lift the restriction, we factorize $\mathbf{W}$ into $\mathbf{W}_1 \times \mathbf{W}_2$, where $\mathbf{W}_1 \in \mathbb{R}^{\mathtt{D} \times \mathtt{F}}$, $\mathbf{W}_2 \in \mathbb{R}^{\mathtt{F} \times \mathtt{D}}$, and $\mathtt{F} (> \mathtt{D})$ is a user-specified hyperparameter. 
Similarly, the bound condition in Equation~\ref{equ:fp_bound} is added during iterations.

As a result, \textsc{MsgProp} possesses the following properties.
\vspace*{-5pt}
\begin{lemma}\label{lemma:fp_to_mp}
The feature propagation (\textsc{FeaProp}) process is a special case of the message propagation (\textsc{MsgProp}) process.
\end{lemma}
\vspace*{-5pt}
\begin{proof}
\textsc{FeaProp} is a special case of \textsc{MsgProp} in which the correlation $c(\mathbf{x}_i^{(l)}, \mathbf{x}_k^{(l)})$ equals $\Tilde{\mathbf{A}}_{ik}$ and $\mathbf{W}_1$ and $\mathbf{W}_2$ are both set to identity matrices. 
\end{proof}
\vspace*{-5pt}
Lemma~\ref{lemma:fp_to_mp} indicates that \textsc{MsgProp} has a higher learning capability than \textsc{FeaProp}.
\vspace*{-5pt}
\begin{lemma}\label{lemma:msgpassing}
The \textsc{MsgProp} process incorporates the classical message passing mechanism~\cite{gilmer2017neural,battaglia2018relational} in graph learning.
\end{lemma}
\begin{proof}
We rewrite Equation~\ref{equ:mp_solution_1} as 
\begin{equation}\label{equ:mp_mp}
\mathbf{z}_i^{(l+1)} = \sum\nolimits_{\mathbf{x}_k \in {\mathcal{N}_i}} c(\mathbf{x}_i^{(l)}, \mathbf{x}_k^{(l)}) \mathbf{x}_k^{(l)} \mathbf{W}_1,
\end{equation}
\begin{equation}\label{equ:mp_solution_2}
\mathbf{\tilde{x}}_i^{(l+1)} = \mathbf{z}_i^{(l+1)} \mathbf{W}_2.
\end{equation}
In particular, computing $\mathbf{z}_i^{(l+1)}$ in Equation~\ref{equ:mp_mp} is a classical message passing process in graph learning. More specifically, Equation~\ref{equ:mp_mp} implements the graph convolution operator~\cite{kipf2016semi} if $c(\mathbf{x}_i^{(l)}, \mathbf{x}_k^{(l)})$ equals $\Tilde{\mathbf{A}}_{ik}$ and implements the graph attention operator~\cite{velivckovic2017graph} if $c(\mathbf{x}_i^{(l)}, \mathbf{x}_k^{(l)})$ is an attention function.
\end{proof}
\vspace*{-5pt}
{Since \textsc{MsgProp} employs message passing, \textsc{MsgProp} is able to utilize existing message passing modules such as the Graph Attention Unit (GAT)~\cite{velivckovic2017graph}, the Graph Convolution Unit (GCN)~\cite{kipf2016semi}, and GraphSAGE~\cite{hamilton2017inductive}, as well as their accompanying optimization techniques.}
Next, we propose the \textsc{MsgProp} imputation network that reconstructs the missing values of the data instances based on the \textsc{MsgProp} process over the similarity graph.

\vspace*{-5pt}
\subsection{\textsc{MsgProp} Imputation Network (MPIN)}
\label{ssec:imputation_network}


We first give the architecture of MPIN and then show how it imputes the missing values of data instances of an input data chunk $\mathcal{X}_a$.

\begin{figure}[!ht]
\centering
\includegraphics[width=0.9\columnwidth]{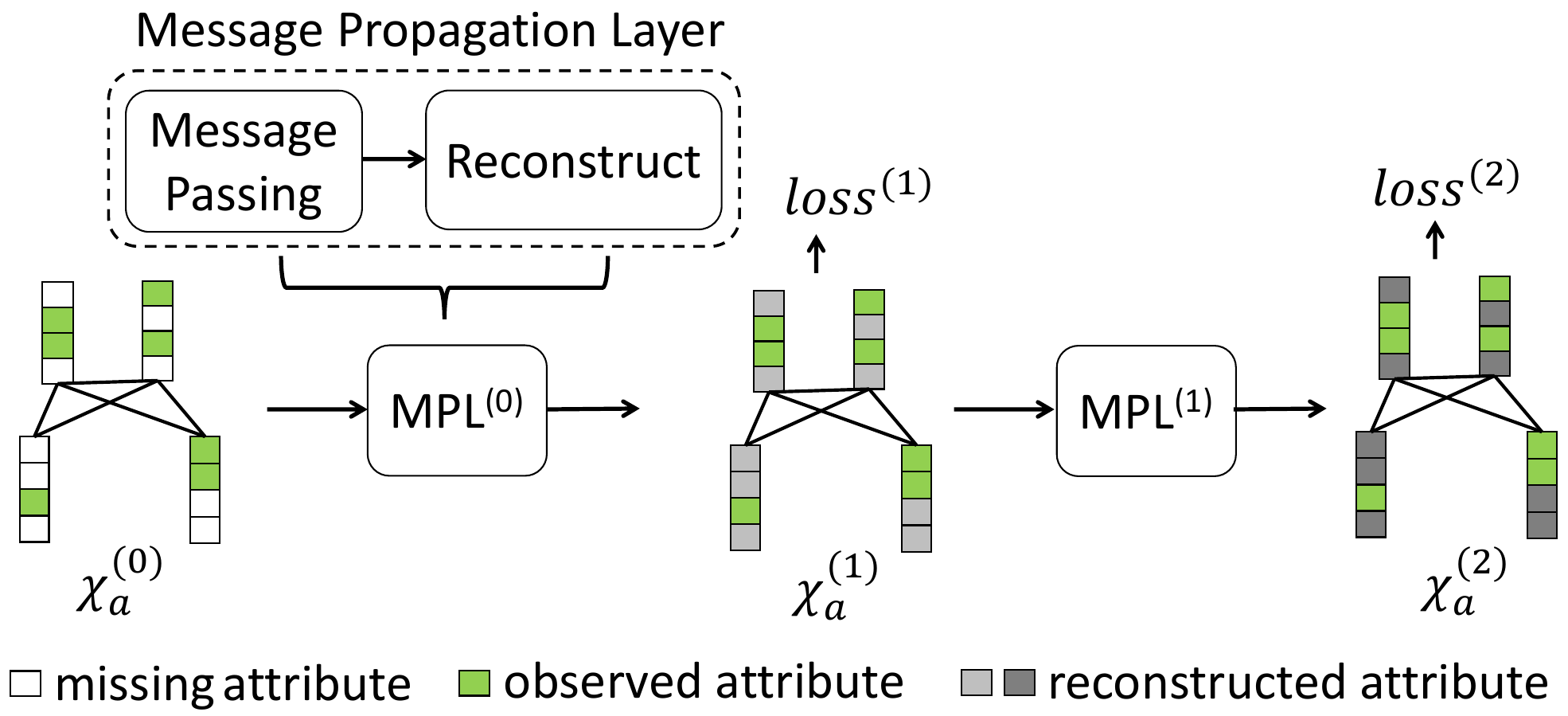}
\caption{Message propagation imputation network.}
\label{fig:MPIN}
\end{figure}

\smallskip
\noindent\textbf{Architecture}. 
MPIN is constructed using a stack of \emph{\textsc{MsgProp} layers} ($\operatorname{MPL}$ for short), a basic internal unit. Specifically, each $\operatorname{MPL}(\cdot)$, consists of two modules, namely the \emph{message passing module} and the \emph{reconstruction module}, as depicted in Figure~\ref{fig:MPIN}.
%

The message passing module corresponds to the computational process presented in Equation~\ref{equ:mp_mp} (see Lemma~\ref{lemma:msgpassing}). The reconstruction module is the combination of a linear transformation and a bound condition, corresponding to the processes in Equations~\ref{equ:mp_solution_2} and~\ref{equ:fp_bound}, respectively.
An $\operatorname{MPL}$ transforms the raw data instance $\mathbf{x}_i^{(0)}$ to $\mathbf{x}_i^{(1)}$ by aggregating the features of its similar nodes in the similarity graph.
This process thus reconstructs the missing values in $\mathbf{x}_i^{(0)}$ by utilizing correlations among the data instances.

The process is iterated by our MPIN to find the optimal reconstruction results for the missing values in the data instances. Typically, we stack two $\operatorname{MPL}$s to build the imputation network MPIN:
\begin{equation}\label{equ:mpl}
\mathcal{X}_a^{(1)} = \operatorname{MPL}(\mathcal{X}_a^{(0)}), \\ \,\,\,\,
\mathcal{X}_a^{(2)} = \operatorname{MPL}(\mathcal{X}_a^{(1)}),
\end{equation}
where $\mathcal{X}_a^{(0)}$ denotes the raw input data chunk and  $\mathcal{X}_a^{(1)}$ denotes the reconstructed data chunk after the first {MPL}. Next,   $\mathcal{X}_a^{(2)}$ performs an enhanced reconstruction based on $\mathcal{X}_a^{(1)}$ and is taken as the final reconstructed result of $\mathcal{X}_a$.

\smallskip
\noindent\textbf{Transductive Learning.} 
{Unlike \textsc{FeaProp} that performs matrix multiplications iteratively until convergence, MPIN adopts a transductive learning paradigm~\cite{vapnik2006estimation}. Specifically, given an input data chunk, MPIN imputes the missing values in the input based on backpropagation of the reconstruction loss corresponding to the observed values (see Equation~\ref{equ:loss_func}).
In other words, there are no separate phases of training and inference, and training data (i.e., observed values) and testing data (missing values) are all included in the training process. This paradigm has been proven effective at imputing missing values effectively~\cite{cao2018brits,du2023saits,andrea2021filling}. The intuition is that if the reconstructed results corresponding to the observed values are close to the observed values themselves, then the reconstructed results corresponding to the missing values should also be close to their ground-truth values though these are unknown in reality.
Accordingly, the training loss of MPIN is defined based on the reconstruction error between the observed values in $\mathcal{X}_a$ and their corresponding values as imputed by {\MPIN{}}.}

Since \MPIN{} generates multiple reconstruction results recursively (see Equation~\ref{equ:mpl}), we inform \MPIN{} about the reconstruction error of each reconstruction process instead of only at the last one. Thus, we use a linear combination of the losses computed from each of the reconstruction results as the overall loss:
\begin{equation}\label{equ:loss_func}
\begin{aligned}
\mathcal{L}^{(1)} =~ &  \operatorname{MSE}(\mathcal{X}_a^{(0)} \odot \mathcal{M}_a, \mathcal{\Tilde{X}}_a^{(1)} \odot \mathcal{M}_a), \\
\mathcal{L}^{(2)} =~ &  \operatorname{MSE}(\mathcal{X}_a^{(0)} \odot \mathcal{M}_a, \Tilde{\mathcal{X}}_a^{(2)} \odot \mathcal{M}_a), \\
\mathcal{L} =~ & \lambda_1 \cdot \mathcal{L}^{(1)} +\lambda_2 \cdot \mathcal{L}^{(2)},
\end{aligned}
\vspace*{-5pt}
\end{equation}
where $\operatorname{MSE}(\cdot, \cdot)$ calculates the mean squared error and $\mathcal{L}^{(1)}$ and $\mathcal{L}^{(2)}$ refer to the reconstruction loss of the first and second \textsc{MsgProp} layer, respectively. Further, $\mathcal{X}_a^{(0)} \odot \mathcal{M}_a$ is a masking process that picks up only the observed values of the data chunk $\mathcal{X}_a^{(0)}$. Also, $\mathcal{\Tilde{X}}_a^{(1)}$, instead of $\mathcal{{X}}_a^{(1)}$, is used when computing the loss since the imputed results of the observed values exist only in $\mathcal{\Tilde{X}}_a^{(1)}$ (see  Equation~\ref{equ:mp_solution_2}), which is an imputation result before the bound condition (i.e., Equation~\ref{equ:fp_bound}) is applied.  Finally, $\lambda_1$ and $\lambda_2$ are hyperparameters.

\begin{algorithm}
\small
\caption{\textsc{Training for Imputation} (input data chunk $\mathcal{X}_a$, training epochs $P$, validation ratio $\beta$, model state $\theta'$)} \label{alg:impute_MPIN}
\begin{algorithmic}[1]

\State validation dataset $\Phi_a \gets $  fraction $\beta $ of observed values in $\mathcal{X}_a$ 
\State $\mathcal{X}'_a \gets
 \mathcal{X}_a \backslash \Phi_a$
\State initialization: $\mathtt{MAE}^{\ast} \gets  \infty$; $\mathcal{X}^{\ast}_a \gets  \mathcal{X}_a$; ${\theta}^{\ast}_a \gets \theta'$
\State initialize \MPIN{} $\gets$ {\MPIN{}} of state ${\theta'}$

\For {p = 1 to P}
    \State $\hat{\mathcal{X}}_a, \hat{\Phi}_a, \hat{\theta}_a \gets$ one-pass backpropagation with \MPIN{}, $\mathcal{X}'_a$

    \State  $\mathtt{MAE}$ $\gets$ MAE ($\hat{\Phi}_a$, $\Phi_a$)\Comment{measuring validation error}
    \If {$\mathtt{MAE} \leq \mathtt{MAE}^{\ast}$ }
        \State  $\mathtt{MAE}^{\ast} \gets  \mathtt{MAE}$; $\mathcal{X}^{\ast}_a \gets  \hat{\mathcal{X}}_a$ ;  ${\theta}^{\ast}_a \gets \hat{\theta}_a $\Comment{update optimal}
    \EndIf
\EndFor
\State $\mathcal{X}^{\ast}_a \gets $  $\mathcal{X}^{\ast}_a \oplus \Phi_a$ \Comment{recover manually removed values}

\State return $\mathcal{X}^{\ast}_a$,  ${\theta}^{\ast}_a$
\end{algorithmic}
\end{algorithm}

Algorithm~\ref{alg:impute_MPIN} shows how to impute missing values of $\mathcal{X}_a$ by training \MPIN{}. First, we generate a validation dataset by randomly removing a fraction (e.g., 5\%) of the observed values in $\mathcal{X}_a$ and use the removed values as the ground truth values to validate the imputation error in each iteration of training (lines 1--2). Specifically, 
we enter the pre-processed data chunk $\mathcal{X}'_a$ into \MPIN{}. Through backpropagation of reconstruction loss (see Equation~\ref{equ:loss_func}), we obtain an imputed data chunk $\mathcal{\hat{X}}_a$ that contains the imputed results of the validation part, i.e., $\hat{\Phi}_a$, and a state of the model (i.e., parameters) denoted by $\hat{\theta}_a$ (line~6). By measuring the mean absolute error (MAE) on the validation dataset, we can find the optimal results of imputation and the optimal state of the model during the iterative training (lines 7--9). Finally, we recover the manually removed original values in $\mathcal{X}_a^{\ast}$, and return it as the final reconstructed result (lines 10-11). The optimal model parameters are returned as well. Their use will be detailed in Section~\ref{ssec:model_update}. 


\vspace*{-5pt}
\subsection{Discussion}\label{ssec:discussion}

The {MPIN} model is inspired by a recent study~\cite{rossi2022unreasonable} that uses feature propagation (\textsc{FeaProp}) to improve the learning effectiveness on graphs with missing values in the node features. 
{Nevertheless, MPIN differs from \textsc{FeaProp} in three key aspects.
First, our imputation process is based on the novel \textsc{MsgProp} process that exploits a correlation function instead of an adjacency matrix to capture correlations among nodes (i.e., instances). Second, \textsc{MsgProp} utilizes the correlations among the attributes for imputation, whereas \textsc{FeaProp} does not. As we have shown, \textsc{FeaProp} is a special case of  \textsc{MsgProp}. 
Third, we reconstruct the missing values in node features via transductive learning based on observed values in the node features. In contrast, \textsc{FeaProp} does not involve learning but simply imputes missing values in node features based on iterative adjacency matrix multiplications.} 
The more comprehensive design makes our approach more effective, as to be shown in Section~\ref{sec:experiments}. {We also give an example below to illustrate the difference between \textsc{FeaProp} and \textsc{MsgProp} at imputation. For simplicity, we only show one of their iterative processes. 
\vspace*{-5pt}
\begin{example}
   Suppose that  a data chunk $\mathcal{X}_1$in the time window $[\mathtt{T} : 2\mathtt{T}]$ contains data instances $\mathbf{x}_2 = [12, 5, 6]$, $\mathbf{x}_3 = [8, 5, 3]$, and $\mathbf{x}_4 = [3, 2, 1]$. If we use $K=2$ to build the similarity graph, both $\mathbf{x}_3$ and $\mathbf{x}_4$ are adjacent nodes to $\mathbf{x}_2$ in the graph. For some reason (e.g., a sensor error), $\mathbf{x}_2$ misses its first component and becomes $\mathbf{x}_2=[null, 5, 6]$. In order to impute the $null$ value in $\mathbf{x}_2$, \textsc{FeaProp} multiplies the normalized adjacency matrix with neighboring instances (see Equation~\ref{equ:fp_solution}) and obtains an imputed result $\mathbf{\tilde{x}}_2 = 0.5\cdot \mathbf{x}_3 + 0.5\cdot \mathbf{x}_4 = [5.5, 3.5, 2]$. With the bound condition, $\mathbf{\tilde{x}}_2$ becomes $[5.5, 5, 6]$. Thus, the $null$ value in $\mathbf{x}_2$ is replaced by $5.5$. 
   In contrast, instead of adopting the equal weight (i.e., 0.5) from the normalized adjacency matrix, \textsc{MsgProp} learns the correlation between $\mathbf{x}_2$ and its neighboring nodes through transductive learning based on the observed values. For simplicity, the learning process is simulated by our observation that finds $\mathbf{x}_3$ to be much closer to $\mathbf{x}_2$ than $\mathbf{x}_4$. Hence, we assume the following correlation values: $\mathbf{c}(\mathbf{x}_2, \mathbf{x}_3)=0.8$ and $\mathbf{c}(\mathbf{x}_2, \mathbf{x}_4)=0.2$. Applying \textsc{MsgProp} (see Equation~\ref{equ:mp_solution_1})~\footnote{For simplicity, we skip the discussion of the correlation among attributes in the example and assume that $\mathbf{W}$ in Equation~\ref{equ:mp_solution_1} is an identity matrix.}, we get $\mathbf{\tilde{x}}_2 = 0.8\cdot \mathbf{x}_3 + 0.2\cdot \mathbf{x}_4 = [7.0, 4.4, 2.6]$. Due to the same bound condition, $\mathbf{x}_2$ becomes $[7.0, 5, 6]$. As $7.0$ is closer to the ground truth (i.e., $12$) than $5.5$, \textsc{MsgProp} is more effective at imputation than \textsc{FeaProp}.
\end{example}}
\vspace*{-5pt}
\section{Continuous Imputation}
\label{sec:incremental}

\subsection{Motivation and Overview}
\label{ssec:incre_impute_pro}

\begin{figure}[!ht]
\centering
\includegraphics[width=0.9\columnwidth]{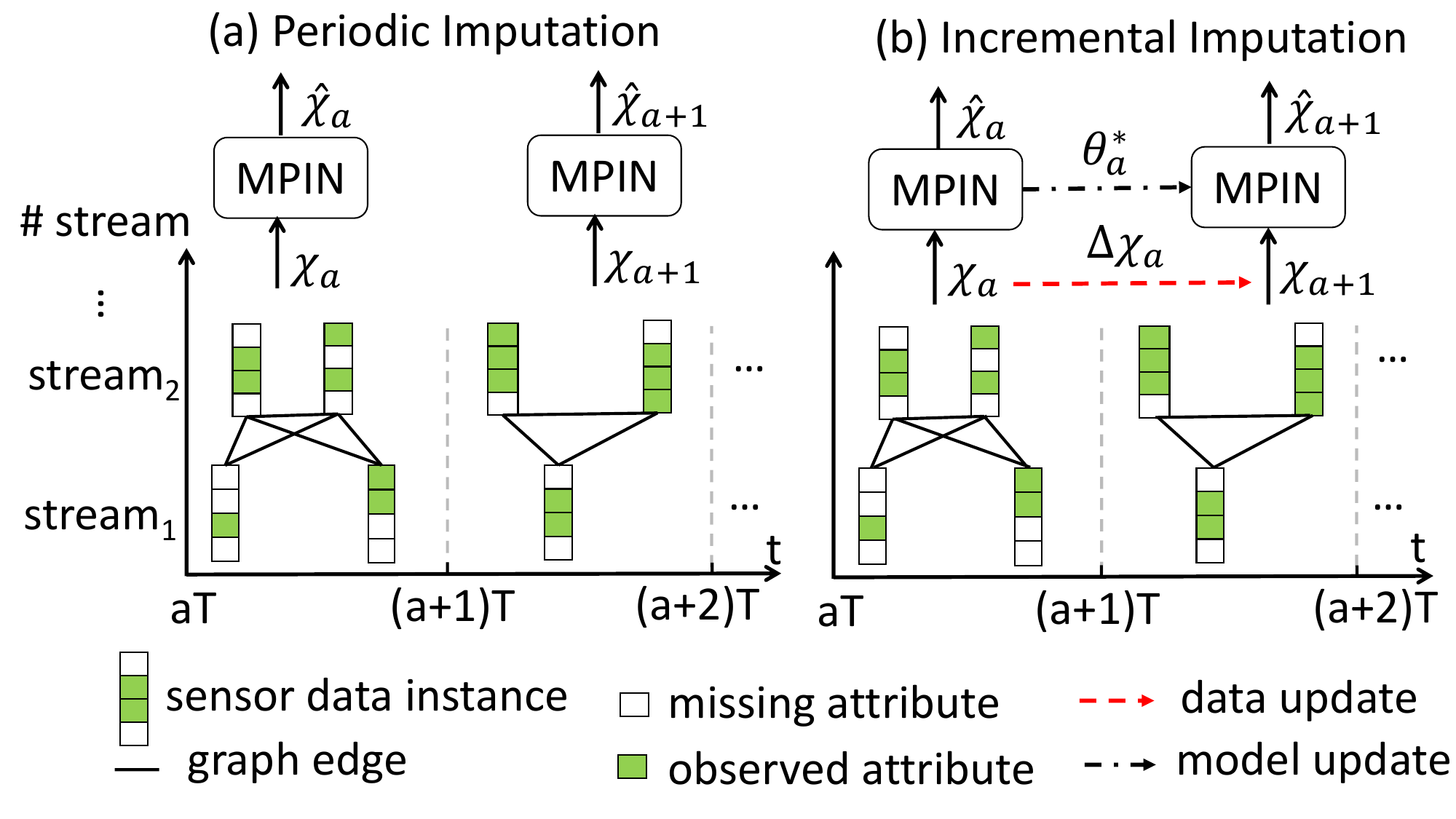}
\caption{Continuous imputation for sensor data streams.}
\label{fig:incre_impute}
\vspace*{-5pt}
\end{figure}

To enable continuous imputation on unbounded sensor data streams, a straightforward approach is to apply {MPIN} to each time window periodically. This approach is called {MPIN-P} and is depicted in Figure~\ref{fig:incre_impute} (a). However, this approach suffers from two significant drawbacks.
First, from a \textbf{data perspective}, the varying number of instances in a data chunk may not provide sufficient data for training {MPIN} effectively at each time window, leading to potentially poor imputation results. Further, earlier data instances in the stream contain valuable information that can improve current and future imputation outcomes, thus making it unwise to disregard these. Second, from a \textbf{model perspective}, since MPIN relies on transductive learning, the entire imputation process (i.e., Algorithm~\ref{alg:impute_MPIN}) must be started from scratch for each window, which is time-consuming, especially for large-sized data chunks.

To address these drawbacks, we propose an incremental framework for continuously imputing sensor data streams.
As shown in Figure~\ref{fig:incre_impute} (b), the incremental imputation framework is powered by the \textbf{data update} and \textbf{model update} mechanisms, corresponding to the data and model challenges. 
%
First, instead of caching all previous data instances for training the current MPIN, the data update mechanism selects and caches the most valuable data instances to enable a time- and space-efficient transductive learning process during the continuous imputation.
Second, to avoid starting from scratch at each window, the model update mechanism regards the model trained at the previous window as a pre-trained model and fine-tunes it for imputing the current data chunk in a transfer learning manner.
We use MPIN-DM to refer to our incremental framework.
%

We proceed to elaborate on the two update mechanisms in Sections~\ref{ssec:data_update} and~\ref{ssec:model_update}, followed by a complexity analysis of the MPIN-DM framework in Sections~\ref{ssec:analysis}.

\subsection{Data Update Mechanism}\label{ssec:data_update}

\begin{figure}[!ht]
\centering
\includegraphics[width=0.7\columnwidth]{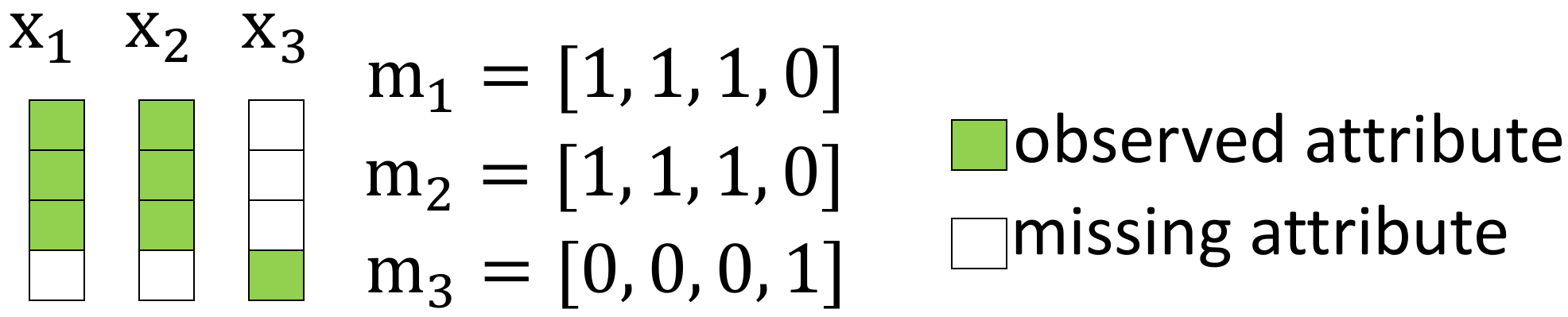}
\caption{{Example of the commonly observed value ratio.}}
\label{fig:sample_score_exa}
\vspace*{-5pt}
\end{figure}

\noindent\textbf{Data Instance Importance Scores.} The data update mechanism aims to identify and retain only the most valuable historical data instances that can help impute the missing values in the current data chunk.
We proceed to formulate criteria for quantifying the value of a data instance.
\vspace*{-5pt}
\begin{criterion}[Higher Observation Ratio]\label{criterion:high_observation}
Intuitively, a sensor data instance with a higher fraction of observed values can contribute more to the imputation of missing values of other data instances.
\end{criterion}
\vspace*{-5pt}
In MPIN, the learning is driven by minimizing the difference between the observed values and their reconstructed counterparts.
In this sense, more observed values can improve the effectiveness of transductive learning.
\vspace*{-5pt}
\begin{criterion}[Lower Observation Overlap Ratio]\label{criterion:less_common_observation}
A sensor data instance whose observed dimensions overlap less with those of other instances can contribute more to the imputation of missing values of other instances.
\end{criterion}
\vspace*{-5pt}
{Generally speaking, the observed values of an instance can provide information for imputing the missing values of other instances only if the former are located in different dimensions than the latter. Referring to the example in Figure~\ref{fig:sample_score_exa}, suppose that both $\mathbf{x}_2$ and $\mathbf{x}_3$ are used to help impute missing values in $\mathbf{x}_1$. Although $\mathbf{x}_2$ has a higher observation ratio, $\mathbf{x}_3$ may be more helpful in imputing $\mathbf{x}_1$'s missing value since all the observed values of $\mathbf{x}_2$ concern the same dimensions as those of $\mathbf{x}_1$ and thus provide little information to the imputation of $\mathbf{x}_1$'s missing value.}

Combining the two criteria, we compute the \textbf{importance score} of a data instance $\mathbf{x}_i$, denoted as $\varphi(\mathbf{x}_i)$, as follows.
\begin{equation}\label{equ:value}
\varphi(\mathbf{x}_i) = \mathit{OR}(\mathbf{x}_i) -  \frac{1}{|V|-1} \sum\nolimits_{\mathbf{x}_k \in V \setminus \mathbf{x}_i} \mathit{OOR}(\mathbf{x}_i, \mathbf{x}_k),
\end{equation}
\begin{equation}\label{equ:hor}
\mathit{OR}(\mathbf{x}_i) = ||\mathbf{m}_i||_0 = \mathbf{m}^\mathsf{T}_i \mathbf{m}_i,
\end{equation}
\begin{equation}\label{equ:lco}
\mathit{OOR}(\mathbf{x}_i, \mathbf{x}_k) = ||\mathbf{m}_i \cap \mathbf{m}_k||_0 = \mathbf{m}^\mathsf{T}_i \mathbf{m}_k.
\end{equation}

In particular, the importance score $\varphi(\mathbf{x}_i)$ in Equation~\ref{equ:value} equals the observation ratio score $\mathit{OR}(\mathbf{x}_i)$ minus the observation overlap ratio score $\mathit{OOR}(\mathbf{x}_i, \mathbf{x}_k)$ averaged over all other instances.
Corresponding to Criterion~\ref{criterion:high_observation}, Equation~\ref{equ:hor} calculates $\mathit{OR}(\mathbf{x}_i)$ as the number of non-zero elements (i.e., $||\cdot||_0$) in the corresponding mask $\mathbf{m}_i$.
Corresponding to Criterion~\ref{criterion:less_common_observation}, Equation~\ref{equ:lco} calculates $\mathit{OOR}(\mathbf{x}_i, \mathbf{x}_k)$ for two instances as the number of non-zero elements in the element-wise AND result of their masks $\mathbf{m}_i$ and $\mathbf{m}_k$.
In both Equations~\ref{equ:hor} and~\ref{equ:lco}, we use the number of non-zero elements instead of the ratio, observing that the feature dimension $\mathtt{D}$ is constant.

The importance score has the following important property.
\begin{lemma}\label{lemma:value_range} 
$\forall \mathbf{x}_i \in V$, we have $0 \leq \varphi(\mathbf{x}_i) \leq \mathtt{D} - 1$.
\end{lemma}
\vspace*{-5pt}






The proof is given in Section~\ref{ssec:proof_lem3} in the Appendix for the sake of brevity.
Lemma~\ref{lemma:value_range} gives the range of the importance score of any data instance.
Based on Lemma~\ref{lemma:value_range}, we are able to normalize the importance scores by dividing them by $(\mathtt{D}-1)$ and use a threshold $\eta \in (0,1]$ to drop those data instances whose importance score does not exceed $\eta$. In practice, $\eta$ is tuned to a proper value based on preliminary experiments, and a value of around 0.6 has been demonstrated to be able to obtain satisfactory performance, as reported in Section~\ref{ssec:effect_value_thre} in the Appendix.

\noindent\textbf{Efficient Computing of Importance Scores}.
It is time-consuming to compute the importance score per data instance.
To accelerate the computations, we 
use matrix multiplication to obtain importance scores of all instances in one pass.
Specifically, we introduce a \textbf{gram mask matrix} as follows.
First, recall that the similarity graph is built based on the input data chunk, i.e., $\mathcal{X}_a \in \mathbb{R}^{|V|\times \mathtt{D}}$, and its corresponding mask chunk is denoted as $\mathcal{M}_a \in \mathbb{R}^{|V|\times \mathtt{D}}$.
Thus, a gram mask matrix $\mathbf{M}^\text{GM}$ is computed as $\mathcal{M}_a \mathcal{M}_a^\textsf{T}$ with size $|V| \times |V|$.
Given the gram mask matrix $\mathbf{M}^\text{GM}$, the importance scores of all data instances can be computed in one pass via the following lemma.

\begin{lemma}\label{lemma:gram_computation}
Given the corresponding gram mask matrix $\mathbf{M}^\text{GM}$, the importance scores  of data instances in set $V$ can be computed jointly by
\begin{equation}\label{equ:sample_score_matrix}
    \boldsymbol{\varphi} = \frac{(|V| * \operatorname{diag}^{-1}(\mathbf{M}^\text{GM}) -  \mathbf{M}^\text{GM} \cdot \mathbf{1}^{|V| \times 1})}{|V|-1},
\end{equation}
where $\boldsymbol{\varphi} \in \mathbb{R}^{|V| \times 1}$ and $\boldsymbol{\varphi}[i]$ captures the importance score of the $i$-th data instance $\mathbf{x}_i$ in data chunk $\mathcal{X}_a$, $\operatorname{diag}^{-1}(\cdot)$ gets the diagonal vector from the input matrix, and $*$ is the element-wise scalar product.
The proof is in Section~\ref{ssec:proof_lem4} in the Appendix for the sake of brevity.
\end{lemma}


\noindent\textbf{Algorithm}. Using Lemma~\ref{lemma:gram_computation}, we can obtain the importance scores of all instances in parallel with GPU-based matrix multiplication.
The mechanism is formalized in Algorithm~\ref{alg:data_update}. First, the cached valuable data instances so far, i.e., $\boldsymbol\Delta\mathcal{X}_{a}$, are concatenated with the current input data chunk to form a new data chunk $\mathcal{X}''_{a+1}$ (line~1). The new data chunk instead of $\mathcal{X}_{a+1}$ is used to train {MPIN}. Likewise, we can obtain a new mask chunk $\mathcal{M}''_{a+1}$ (line~1), based on which we calculate the gram mask matrix and further exploit Lemma 2 to derive the importance score of each data instance (lines~2--3). Finally, we keep those with importance scores above the threshold $\eta$ and regard them as the updated valuable instances so far (lines~4--5).

\begin{algorithm}
\small
\caption{\textsc{DataUpdate} (last valuable instances $\boldsymbol\Delta\mathcal{X}_{a}$ and mask $\boldsymbol\Delta\mathcal{M}_a$, current data chunk $\mathcal{X}_{a+1}$ and mask $\mathcal{M}_{a+1}$, predefined value threshold $\eta$)} \label{alg:data_update}
\begin{algorithmic}[1]
\State $\mathcal{X}''_{a+1} \gets concat(\boldsymbol\Delta\mathcal{X}_{a},  \mathcal{X}_{a+1})$; $\mathcal{M}''_{a+1} \gets concat({\boldsymbol\Delta\mathcal{M}_a}, \mathcal{M}_{a+1})$
\State the gram mask matrix $\mathbf{M}^\text{GM} \gets \mathcal{M}_{a+1}'' {\mathcal{M}_{a+1}''}^\textsf{T}$
\State compute $\boldsymbol{\varphi}$ with $\mathbf{M}^\text{GM}$ using Equation~\ref{equ:sample_score_matrix}
\State $\boldsymbol\Delta\mathcal{X}_{a+1} \gets \{\mathcal{X}''_{a+1}[i] \mid \boldsymbol{\varphi}[i] \geq \eta \}$
\State \Return $\boldsymbol\Delta\mathcal{X}_{a+1}$
\end{algorithmic}
\end{algorithm}

{To explain the process of data update and its benefits to imputation, we give an example.
\vspace*{-5pt}
\begin{example}
    To continue Example 1, we assume that another time window $[0 : \mathtt{T}]$ is just before the window $[\mathtt{T} : 2\mathtt{T}]$. Suppose that the data chunk $\mathcal{X}_0$ in $[0, \mathtt{T}]$ contains 2 data instances $\mathbf{x}_0 = [null, null, 4]$ and $\mathbf{x}_1 = [12, 5, 5]$. Hence, their mask vectors are $\mathbf{m}_0 = [0, 0, 1]$ and $\mathbf{m}_1 = [1, 1, 1]$, respectively, and the mask matrix is $\mathcal{M}_0 = [\mathbf{m}_0, \mathbf{m}_1]$. The data update strategy enables us to pass valuable instances to the next window. First, we compute the gram mask matrix $\mathbf{M}^{GM} = \mathcal{M}_0\mathcal{M}_0^{\mathtt{T}}= [[1, 1],[1, 3]]$ and the importance score vector $[0, 1]$ based on Equation~\ref{equ:sample_score_matrix}. Thus, the importance scores of $\mathbf{x}_0$ and $\mathbf{x}_1$ are  $0$ and $1$, meaning that $\mathbf{x}_0$ is much less important since it has very few observed attributes. As a result, $\mathbf{x}_1$ is taken to the next window, i.e., $[\mathtt{T} : 2\mathtt{T}]$. 
    Next, we show what difference $\mathbf{x}_1$ can make to the imputation in Example 1. In that example, $\mathbf{x}_1$ will replace $\mathbf{x}_4$ to be the new adjacent node of $\mathbf{x}_2$ in the similarity graph since $\mathbf{x}_1$ is closer to $\mathbf{x}_2$ than $\mathbf{x}_4$. Accordingly, the  correlation values are updated to $\mathbf{c}(\mathbf{x}_2, \mathbf{x}_3)=0.2$ and $\mathbf{c}(\mathbf{x}_2, \mathbf{x}_1)=0.8$, given that $\mathbf{x}_1$ now is closer to $\mathbf{x}_2$ than is $\mathbf{x}_3$. Following the \textsc{MsgProp} process in Example 1, the imputation result is $[11.2, 5, 6]$, i.e., $11.2$ is the imputed value for the $null$ in $\mathbf{x}_2$. This is even closer to the ground truth than the imputed value of $7.0$ in Example 1.
\end{example}}

\vspace*{-5pt}
\subsection{Model Update Mechanism}
\label{ssec:model_update}

To avoid the costly process of training MPIN from scratch for every time window, we propose a model update mechanism that resumes training from the best state so far. The mechanism consists of two components, namely
\emph{model state selection} and \emph{model update}.


\noindent\textbf{{Model State Selection.}}
This component helps choose the best model state, i.e., the best-trained parameters so far. The best state is then used as the initial state of MPIN in the next window.
To make the resumed training of \MPIN{} in the next window effective, we need to ensure a "good" initial state (i.e., parameters) for \MPIN{} from history. We realize this by Algorithm~\ref{alg:model_state_sel}. First, we find the optimal state of the model within the current time window using Algorithm~\ref{alg:impute_MPIN}. This will in turn be conveyed to the next window as the initial state of \MPIN{} for retraining. As we can see, ${\theta}^{\ast}_{a+1}$ is derived from ${\theta}^{\ast}_{a}$, i.e., the best state so far. This way, we can ensure that ${\theta}^{\ast}_{a+1}$ is always the best state of the model seen so far.

\begin{algorithm}
\small
\caption{\textsc{ModelStateSelection} (last best state $\theta^{\ast}_{a}$,  current data chunk $\mathcal{X}_{a+1}$, training epochs P, validation fraction $\eta$)} \label{alg:model_state_sel}
\begin{algorithmic}[1]

\State $\mathcal{X}^{\ast}_{a+1}$,   ${\theta}^{\ast}_{a+1} \gets$ call Algorithm~\ref{alg:impute_MPIN} 
($\mathcal{X}_{a+1}$, P, $\epsilon$, $\theta^{\ast}_{a}$)\Comment{best state} 

\State return ${\theta}^{\ast}_{a+1}$
\end{algorithmic}
\end{algorithm}

\begin{figure}[!ht]
\centering
\includegraphics[width=0.75\columnwidth]{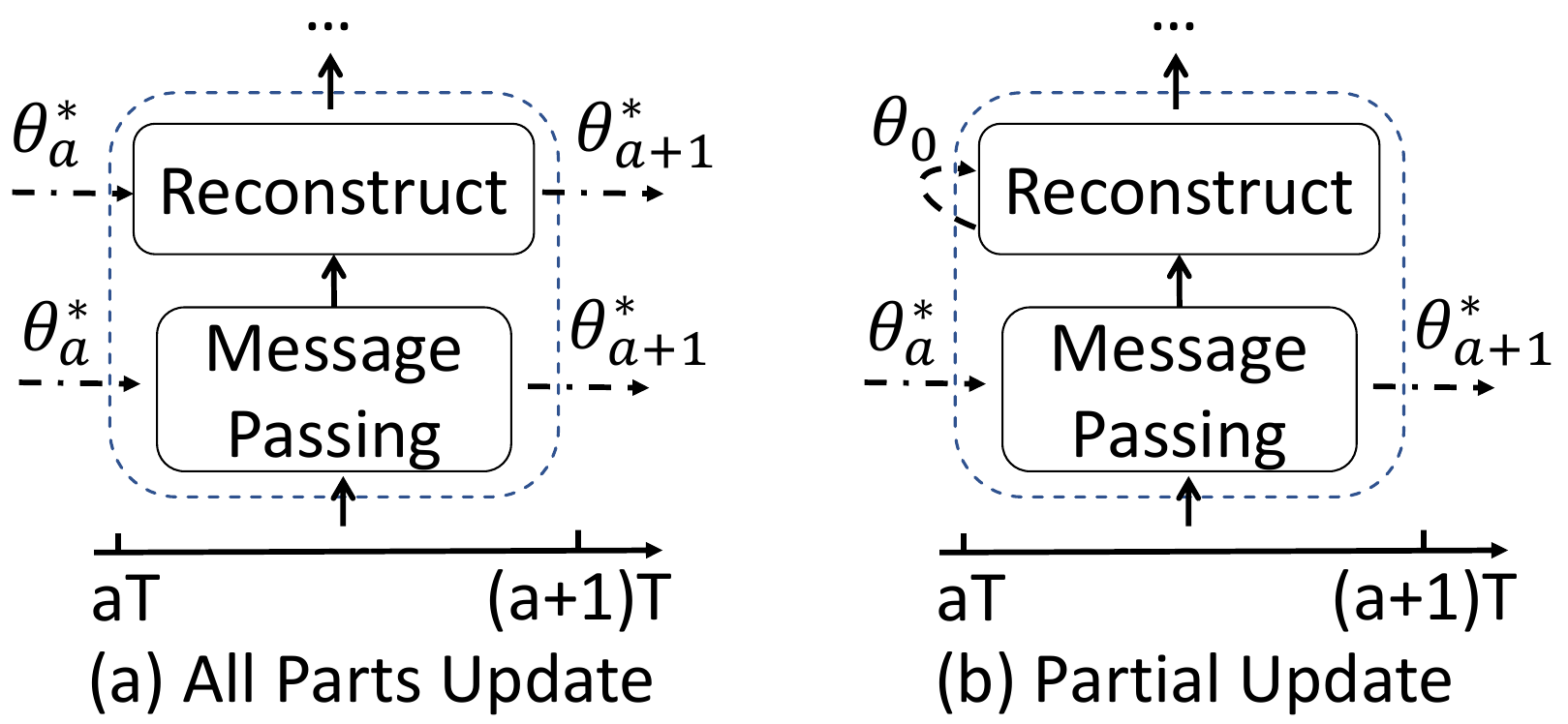}
\caption{All parts update vs. partial update.}
\label{fig:parameter_reuse}
\vspace*{-5pt}
\end{figure}

\noindent\textbf{Model Update.}
After getting the best model state so far, we may simply use it as the initial state to train the whole MPIN for the current window, i.e., from the best state instead of from scratch~\cite{galke2020incremental} (see Figure~\ref{fig:parameter_reuse} (a)). However, our preliminary experiments show that this yields poor imputation results, probably due to overfitting.

To achieve better results, we adopt another strategy to make use of the best state and to initialize the current MPIN.
Specifically, referring to Figure~\ref{fig:parameter_reuse} (b), we only utilize the best state parameters relevant to MPIN's message passing module. In addition, we always initialize MPIN's reconstruction module with random parameter settings (i.e., $\theta_0$). This contributes to preventing overfitting. 
This approach is inspired by transfer learning, where the best state of a model acts as a pre-trained model on a graph (i.e., a data chunk), and we only need to fine-tune it for another graph corresponding to the current data chunk.
The fine-tuning process in our context refers to retraining only the reconstruction module from scratch while reusing the previous parameters of the message-passing modules. 
This way, the knowledge learned by \MPIN{} from the previous window can be transferred to help impute the data chunk in the current window, while significantly reducing the risk of overfitting. 


\vspace*{-5pt}
\subsection{Complexity Analysis}
\label{ssec:analysis}

We present the space and time complexity of the proposed \DMU{} and the periodic approach, \PM{}.

\noindent\textbf{Space Complexity}. 
The space cost mainly corresponds to the storage of relevant data in the streams for online imputation, including the cached data and the new data in the current window.
We assume that the average size of a data chunk is $\mathtt{V}$, that the maximum cached data via data update strategy is $\mathtt{V}^\mathit{mc}$, and that $a$ is the index of the current time window (i.e., the number of time windows that have passed so far). \DMU{} takes $O(\mathtt{V} + \mathtt{V}^\mathit{mc})$ space. In contrast, \PM{} consumes $O(\mathtt{V})$. Furthermore, if an imputation method caches all historical data for retraining the current MPIN, it consumes $O(a \cdot \mathtt{V})$. 
We can regard $\mathtt{V}$ and $\mathtt{V}^\mathit{mc}$ as constants and then get:
\vspace*{-2pt}
\begin{equation}
    O(\mathtt{V} + \mathtt{V}^\mathit{mc}) \approx O(\mathtt{V}) \approx O(1) \ll O(a \cdot \mathtt{V}) \approx O(a)
\end{equation}
\vspace*{-2pt}
As we can see, \DMU{}'s space complexity is at the same level as that of \PM{}, but \DMU{} is much more effective (see Section~\ref{ssec:eva_framework}). Also, the caching-all approach is infeasible as space consumption increases constantly as streams evolve.

\noindent\textbf{Time Complexity}.
The primary time cost is associated with training \MPIN{} to impute missing values in a data chunk (see Algorithm~\ref{alg:impute_MPIN}). 
While it is hard to perform a detailed time complexity analysis, we can provide some insight into the complexity from the following analysis.
For any chunk $\mathcal{X}_{a+1}$ in the $(a+1)$-st time window, if we train \MPIN{} from scratch until achieving the best-imputed results, the time cost is denoted as $\mathcal{T}(\theta_{a+1}^{\ast})$. 
In contrast, utilizing the model update mechanism, the training resumes from last best state $\theta_{a}^{\ast}$, resulting in a reduced time cost of $\Delta \mathcal{T} = \mathcal{T}(\theta_{a+1}^{\ast}) - \mathcal{T}(\theta_{a}^{\ast})$.
When the size of $\mathcal{X}_{a+1}$ is large, we have $\Delta \mathcal{T} \ll \mathcal{T}(\theta_{a+1}^{\ast})$, as more data typically entail higher time costs if the current \MPIN{} is trained from scratch.

\section{Experimental Studies}
\label{sec:experiments}

Section~\ref{ssec:setups} covers the overall experimental settings.
Subsequently, Sections~\ref{ssec:eva_network} and~\ref{ssec:eva_framework} validate the efficacy of the snapshot imputer MPIN (see Section~\ref{sec:proposed_model}) and the continuous imputation framework MPIN-DM (see Section~\ref{sec:incremental}), respectively.

\vspace*{-5pt}
\subsection{Overall Settings}
\label{ssec:setups}

\noindent\textbf{Datasets}.
We use three datasets from different  scenarios:
\begin{itemize}[leftmargin=*]

\item \textbf{ICU}\footnote{\url{https://physionet.org/challenge/2012/}} consists of 11,988 48-hour time series that record the health condition of patients in Intensive Care Units (ICU) during the initial 48 hours after their admission to ICU. The records are taken hourly and each one contains 37 variables such as temperature, heart rate, and blood pressure.
 
\item \textbf{Airquality}\footnote{\url{https://archive.ics.uci.edu/ml/datasets/Beijing+Multi-Site+Air-Quality+Data}} contains hourly air quality monitoring data from 12 monitoring sites in Beijing for a continuous period of 48 months (1,461 days). Each data instance is comprised of 11 variables such as PM2.5, PM10, and SO2. Since the application focuses on daily air quality, we aggregate the data from all sites to form concurrent streams spanning a single day, resulting in a total of 17,532 concurrent streams (\#days $\times$ \#sites).

\item \textbf{Wi-Fi}\footnote{\url{https://www.kaggle.com/competitions/indoor-location-navigation/data}} is a sequence of Wi-Fi signal records that span one hour at a mall. Each instance in the sequence represents a vector of 671 Wi-Fi Received Signal Strength Indicator (RSSI) values, each from one of 671 Wi-Fi access points. 
The collected RSSI vectors can be used for indoor positioning~\cite{li2023data}.
\end{itemize}

Table~\ref{tab:dataset_des} presents the key characteristics of the datasets. 
Specifically, {\ICU} and {\AIR} feature concurrent streams with longer lengths and heterogeneous attributes in periodic instances, whereas {\WIFI} consists of high-dimensional homogeneous aperiodic sensor data. 
Additionally, {\ICU} and {\WIFI} have originally high data sparsity, whereas {\AIR} is less sparse.
\begin{table}[!ht]
    \centering
    \footnotesize
    \caption{Key characteristics of the datasets.}
    \label{tab:dataset_des}
    \begin{tabular}{l|ccc}
    \toprule
    {Datasets}  & \ICU & \AIR & \WIFI \\ 
    \midrule
    Dimensionality of Data Instances  & 37 & 11 & 671  \\
    Time Length of Streams (hours)  & 48 & 24 & 1   \\
    Number of Concurrent Streams  & 11988 &  17532 & 1  \\
    Regular Sampling & Yes &  Yes  & No   \\
    Heterogeneous Attributes & Yes &  Yes & No  \\
    Original Data Sparsity   & 80.0\% &1.6\% & 85.6\%  \\
    \bottomrule
    \end{tabular}
    \vspace{-5pt}
\end{table}

\smallskip
\noindent\textbf{Implementation}.
The project is primarily in Python 3.8 and tested on a Linux server with 3.20 GHz Intel Core i9 CPU and NVIDIA Geforce P8 GPU with 24.5 GB memory. 
PyTorch 1.8 is used to create the neural network models. 
The datasets, code, and configuration details are accessible online~\cite{codeRepo}.
To construct the similarity graph, we use Euclidean distance-based $K$NN with $K = 10$.
{For MPIN, we use two \textsc{MsgProp} layers and apply  GraphSAGE~\cite{hamilton2017inductive} unit to implement the internal message passing module\footnote{{Other alternative message passing modules such as GAT~\cite{velivckovic2017graph} and GCN~\cite{kipf2016semi} achieve lower performance and are covered in Section~A.6 in Appendix.}}.} 
In model training, we set the learning rate to 0.01, the weight decay rate to 0.1, $\lambda_1$ and $\lambda_2$ to 1, the epoch count to 200, and use Adam as the gradient descent optimizer.
{The Appendix covers additional experiments on the selection of hyperparameters, such as the similarity function, the value K of the KNN-based method for building similarity graphs (see Section~\ref{ssec:similarity_graph}), and the number of \textsc{MsgProp} layers.}

\smallskip
\noindent\textbf{Evaluation Metrics}. We concern both effectiveness and efficiency.

(1) {Effectiveness}.
{As no ground-truth values are available for truly missing values in the real datasets, following previous studies~\cite{cao2018brits,du2023saits,andrea2021filling} on data imputation,
we randomly remove a fraction of the observed attributes in $\mathcal{X}_a$ and use the removed values as the ground truth values to evaluate the imputation error. Specifically,  
we employ \emph{Mean Absolute Error} (MAE) and \emph{Mean Relative Error} (MRE) to measure the difference between the imputed results $\hat{\mathcal{X}_a}$ and the input data chunk $\mathcal{X}_a$ with respect to the randomly removed attribute values that are marked by an indicator matrix $\mathcal{M}_e$. They are calculated as follows.
$$
MAE (\mathcal{X}_a, \hat{\mathcal{X}_a}, \mathcal{M}_e) = \frac{\sum_{i=1}^{\mathtt{J}\cdot\mathtt{T}}\sum_{d=1}^{\mathtt{D}} \big( |\mathcal{X}_a[i, d] - \mathcal{\hat{X}}_a[i, d]| \cdot \mathcal{M}_e[i, d] \big)}{\sum_{i=1}^{\mathtt{J}\cdot \mathtt{T}}\sum_{d=1}^{\mathtt{D}}\mathcal{M}_e[i, d]},
$$
$$
MRE (\mathcal{X}_a, \hat{\mathcal{X}}_a, \mathcal{M}_e) = \frac{\sum_{i=1}^{\mathtt{J}\cdot  \mathtt{T}}\sum_{d=1}^{\mathtt{D}} \big( |\mathcal{X}_a[i, d] - \mathcal{\hat{X}}_a[i, d]| \cdot \mathcal{M}_e[i, d] \big)}{\sum_{i=1}^{\mathtt{J}\cdot \mathtt{T}}\sum_{d=1}^{\mathtt{D}} \big( |\mathcal{X}_a[i, d]| \cdot \mathcal{M}_e[i, d] \big)}.
$$}

The lower values of these metrics indicate higher accuracy in imputed results.
These metrics have been proven effective in assessing the quality of data imputation~\cite{cao2018brits,du2023saits,andrea2021filling}.

(2) {Efficiency}.
As a crucial factor for online imputation, efficiency is reflected in two metrics: \emph{imputation time cost} required to impute a data chunk, and \emph{memory cost} required by the model online. 
For a graph-based imputer, its time cost also includes the construction of a similarity graph before the imputation in a time window.



\vspace*{-5pt}
\subsection{Evaluation on Snapshot Imputation}
\label{ssec:eva_network}

%
In this part, we compare the proposed {MPIN} approach with alternative representative imputers, focusing on data imputation for a single time window. 

\noindent\textbf{Baselines.} We compare our  {MPIN} with seven existing data imputers from three categories, as listed in Table~\ref{tab:baselines}.

\begin{table*}[]
\centering
\small
\caption{Baseline methods.}\label{tab:baselines}
\resizebox{\textwidth}{!}{
\begin{tabular}{@{}c|cl@{}}
\toprule
Category                                                                      & Method & Description                                                                                                                             \\ \midrule
\multirow{4}{*}{Traditional}                                                  & MEAN~\cite{meanImputer}   & MEAN imputes the missing values in a data chunk using the mean of each corresponding dimension.                                              \\ 
                                                                              & KNN~\cite{troyanskaya2001missing}    & KNN imputes the missing values of each instance using the mean of the instance's k nearest neighbors.                                       \\ 
                                                                              & MICE~\cite{azur2011multiple}   & Multiple Imputation by Chained Equation imputes missing values based on interdependent features in a round-robin fashion.        \\ 
                                                                              & MF~\cite{hastie2009elements}     & Matrix Factorization  treats a data chunk as a matrix and employs iterative SVD decomposition to impute missing values.           \\ \midrule
\multirow{2}{*}{\begin{tabular}[c]{@{}c@{}}Time-series \\ based\end{tabular}} & BRITS~\cite{cao2018brits}  & Bi-directional Recurrent Imputation for Time Series data adapts an RNN for iterative imputation.                               \\ 
                                                                              & SAITS~\cite{du2023saits}  & Self-Attention-based Imputation for Time Series  replaces the RNN with the self-attention to capture dependencies for imputation. \\ \midrule
Graph-based                                                                   & FP~\cite{rossi2022unreasonable}     & Feature Propagation captures the correlation of data instances by iterative adjacency matrix multiplications.                              \\ \bottomrule
\end{tabular}
\vspace*{-5pt}}
\end{table*}

\noindent\textbf{Parameter Settings.} Our experiments concern three key parameters. 
%
%
First, we vary the parameter of \emph{missing rate}.  Similar to previous work~\cite{cao2018brits,du2023saits}, we do not have the ground truth for those truly missing values. Therefore, we randomly remove a ratio of observed values from the raw data. The removal ratio is called the missing rate and the removed values are used as the ground truth of the missing values thus induced. 
Second, we vary the parameter of \emph{window length}, i.e., changing the length of a time window. A longer time window will involve more data instances of the stream, but the number of windows will be less due to the fixed stream length. 
Third, we vary the parameter of \emph{concurrent streams}, i.e., the number of concurrent streams involved in a time window. We achieve this by varying the ratio of the original number of streams that are included in the data for our experiments. In general, the lower the ratio, the fewer streams are involved and the smaller a data chunk will be. 

Table~\ref{tab:sec3_para} gives the parameters settings with defaults shown in bold.
%
{It is noteworthy that we apply relatively higher missing rates (up to 80\%)  to {Airquality} because this dataset originally is much less sparse than the other two (see Table~\ref{tab:dataset_des}).} Besides, \WIFI{} has only one single stream and thus there is no variation on the concurrent streams. The parameter of window length is set properly according to the total length of the streams. 
In particular, the length of time windows for ICU and Airquality are at an hourly level, since the data instances in the streams are sampled each hour. However, the actual data amount within a time window is large due to the existence of concurrent streams. Nevertheless, \MPIN{}'s imputation needs a few seconds only (see Table~\ref{tab:overall_time}).

     
      
      

\begin{table}[!htbp]
\centering
\small
\caption{Parameter settings (defaults are in bold).}
\label{tab:sec3_para}
\resizebox{\columnwidth}{!}{
\begin{tabular}{l|cccc}
\toprule
Datasets & ICU & Airquality & Wi-Fi \\
\midrule
Missing Rate (\%) & [10, 30, \textbf{50}] & [40, 60, \textbf{80}] & [10, 30, \textbf{50}] \\
Window Length & [1,2,3,\textbf{4},5,6] (h.) & [1,\textbf{2},3,4,5] (h.) & [2,4,\textbf{6},8,10] (min.) \\
Concurrent Streams (\%) & \multicolumn{2}{c}{[1, 10, 20, $\ldots$, \textbf{100}]} & -- \\
\bottomrule
\end{tabular}}
\vspace*{-5pt}
\end{table}

\vspace*{-5pt}
\subsubsection{Overall Effectiveness Comparison}

We vary the missing rate per dataset to test the robustness of each method under varied data sparsity.
The effectiveness measures are reported in Table~\ref{tab:overall_effective}, where the \best{best} and \secBest{second-best} results per setting are highlighted.

Overall, {MPIN}  always outperforms the competitors with wide margins in terms of both MAE and MRE in almost all settings. Only in very few cases, {MPIN} is the second-best, with performance very close to the best. 
Time series models {SAITS} and {BRITS} can only exploit temporal dependencies in a series (i.e., a stream), whereas {MPIN} is able to exploit correlations among data instances that may include those across different streams. 
Moreover, {MPIN} operates on a graph and thus can exploit positive relational inductive biases from the graph~\cite{battaglia2018relational} to further promote the effectiveness of imputation. This also explains why {MPIN} outperforms those traditional imputers such as MICE and MF. 

{In addition, compared to {FP}, {MPIN} is much more effective in most cases. Unlike \textsc{FeaProp},  the \textsc{MsgProp} mechanism can capture inter-instance and inter-attribute correlations dynamically. Also, we notice that {FP} performs relatively better on \WIFI{} than on \ICU{} and \AIR{}. This is because \WIFI{} is a homogeneous dataset where it is easier for the \textsc{FeaProp} in {FP} to capture correlations.} 

{Furthermore, we find that {MPIN} is much more robust to increasing dataset sparsity than the other methods. 
On the one hand, a similarity graph-based data structure involves data instances across streams and timestamps within a window and thus is more likely to find alternative neighbors when the original instances become sparse. On the other hand, {MPIN} can exploit the correlations in the graph-structured data sufficiently using \textsc{MsgProp} and transductive learning.}


\begin{table*}[]
\small
\caption{Overall effectiveness comparison (the unit of MRE is \%).}\label{tab:overall_effective}
\resizebox{\textwidth}{!}{
\begin{tabular}
{@{}c|cccccc|cccccc|cccccc@{}}
\toprule

Dataset & \multicolumn{6}{c|}{ICU}                                                                                                                                     & \multicolumn{6}{c|}{Airquality}                                                                                                                              & \multicolumn{6}{c}{Wi-Fi}                                                                                                                                     \\ \cmidrule(r){1-1} \cmidrule(lr){2-7} \cmidrule(lr){8-13} \cmidrule(l){14-19}

Rate  & \multicolumn{2}{c|}{50\%}                                 & \multicolumn{2}{c|}{30\%}                                 & \multicolumn{2}{c|}{10\%}            & \multicolumn{2}{c|}{80\%}                                 & \multicolumn{2}{c|}{60\%}                                 & \multicolumn{2}{c|}{40\%}            & \multicolumn{2}{c|}{50\%}                                 & \multicolumn{2}{c|}{30\%}                                 & \multicolumn{2}{c}{10\%}            \\ 
Metrics  & \multicolumn{1}{c|}{MAE}  & \multicolumn{1}{c|}{MRE} & \multicolumn{1}{c|}{MAE}  & \multicolumn{1}{c|}{MRE} & \multicolumn{1}{c|}{MAE}  & MRE  & \multicolumn{1}{c|}{MAE}  & \multicolumn{1}{c|}{MRE } & \multicolumn{1}{c|}{MAE}  & \multicolumn{1}{c|}{MRE } & \multicolumn{1}{c|}{MAE}  & MRE  & \multicolumn{1}{c|}{MAE}  & \multicolumn{1}{c|}{MRE} & \multicolumn{1}{c|}{MAE}  & \multicolumn{1}{c|}{MRE} & \multicolumn{1}{c|}{MAE}  & MRE \\ \midrule
MEAN     & \multicolumn{1}{c|}{0.88} & \multicolumn{1}{c|}{101.7}    & \multicolumn{1}{c|}{0.81} & \multicolumn{1}{c|}{94.98}    & \multicolumn{1}{c|}{0.73} & 84.12    & \multicolumn{1}{c|}{0.48} & \multicolumn{1}{c|}{84.17}    & \multicolumn{1}{c|}{0.46} & \multicolumn{1}{c|}{79.56}    & \multicolumn{1}{c|}{0.40} & 68.98    & \multicolumn{1}{c|}{2.21} & \multicolumn{1}{c|}{92.06}    & \multicolumn{1}{c|}{1.95} & \multicolumn{1}{c|}{81.39}    & \multicolumn{1}{c|}{1.52} & 63.21    \\ 
KNN      & \multicolumn{1}{c|}{0.74} & \multicolumn{1}{c|}{85.26}    & \multicolumn{1}{c|}{0.62} & \multicolumn{1}{c|}{72.76}    & \multicolumn{1}{c|}{0.46} & 53.33    & \multicolumn{1}{c|}{0.48} & \multicolumn{1}{c|}{84.43}    & \multicolumn{1}{c|}{0.47} & \multicolumn{1}{c|}{80.22}    & \multicolumn{1}{c|}{0.41} & 70.44    & \multicolumn{1}{c|}{1.84} & \multicolumn{1}{c|}{76.89}    & \multicolumn{1}{c|}{1.43} & \multicolumn{1}{c|}{59.51}    & \multicolumn{1}{c|}{0.63} & 26.37    \\ 
MICE     & \multicolumn{1}{c|}{0.73} & \multicolumn{1}{c|}{84.85}    & \multicolumn{1}{c|}{0.61} & \multicolumn{1}{c|}{71.55}    & \multicolumn{1}{c|}{0.44} & \secBest{51.07}    & \multicolumn{1}{c|}{0.48} & \multicolumn{1}{c|}{83.52}    & \multicolumn{1}{c|}{0.45} & \multicolumn{1}{c|}{78.02}    & \multicolumn{1}{c|}{0.39} & 66.74    & \multicolumn{1}{c|}{2.00} & \multicolumn{1}{c|}{83.68}    & \multicolumn{1}{c|}{1.65} & \multicolumn{1}{c|}{68.77}    & \multicolumn{1}{c|}{0.99} & 41.25    \\ 
MF       & \multicolumn{1}{c|}{0.79} & \multicolumn{1}{c|}{91.81}    & \multicolumn{1}{c|}{0.71} & \multicolumn{1}{c|}{83.29}    & \multicolumn{1}{c|}{0.57} & 66.37    & \multicolumn{1}{c|}{0.54} & \multicolumn{1}{c|}{91.57}    & \multicolumn{1}{c|}{0.51} & \multicolumn{1}{c|}{86.76}    & \multicolumn{1}{c|}{0.46} & 79.58    & \multicolumn{1}{c|}{1.85} & \multicolumn{1}{c|}{77.19}    & \multicolumn{1}{c|}{1.50} & \multicolumn{1}{c|}{62.34}    & \multicolumn{1}{c|}{0.81} & 33.66    \\ 
FP       & \multicolumn{1}{c|}{0.83} & \multicolumn{1}{c|}{96.03}    & \multicolumn{1}{c|}{0.78} & \multicolumn{1}{c|}{91.49}    & \multicolumn{1}{c|}{0.70} & 81.35    & \multicolumn{1}{c|}{0.57} & \multicolumn{1}{c|}{99.78}    & \multicolumn{1}{c|}{0.57} & \multicolumn{1}{c|}{98.67}    & \multicolumn{1}{c|}{0.56} & 96.36    & \multicolumn{1}{c|}{1.26} & \multicolumn{1}{c|}{52.77}    & \multicolumn{1}{c|}{0.55} & \multicolumn{1}{c|}{\secBest{22.73}}    & \multicolumn{1}{c|}{\best{0.12}} & \best{5.17}     \\ 
BRITS    & \multicolumn{1}{c|}{0.57} & \multicolumn{1}{c|}{70.76}    & \multicolumn{1}{c|}{0.50} & \multicolumn{1}{c|}{63.04}    & \multicolumn{1}{c|}{0.44} & 55.19    & \multicolumn{1}{c|}{0.49} & \multicolumn{1}{c|}{68.64}    & \multicolumn{1}{c|}{0.39} & \multicolumn{1}{c|}{55.57}    & \multicolumn{1}{c|}{0.35} & 49.00    & \multicolumn{1}{c|}{\secBest{0.37}} & \multicolumn{1}{c|}{\secBest{48.28}}    & \multicolumn{1}{c|}{\secBest{0.33}} & \multicolumn{1}{c|}{43.24}    & \multicolumn{1}{c|}{0.30} & 39.69    \\ 
SAITS    & \multicolumn{1}{c|}{\secBest{0.53}} & \multicolumn{1}{c|}{\secBest{65.89}}    & \multicolumn{1}{c|}{\secBest{0.47}} & \multicolumn{1}{c|}{\secBest{58.29}}    & \multicolumn{1}{c|}{\secBest{0.41}} & 
51.63    & \multicolumn{1}{c|}{\secBest{0.43}} & \multicolumn{1}{c|}{\secBest{60.65}}    & \multicolumn{1}{c|}{\secBest{0.31}} & \multicolumn{1}{c|}{\secBest{43.42}}    & \multicolumn{1}{c|}{\secBest{0.23}} & \best{32.69}    & \multicolumn{1}{c|}{0.47} & \multicolumn{1}{c|}{60.92}    & \multicolumn{1}{c|}{0.40} & \multicolumn{1}{c|}{52.04}    & \multicolumn{1}{c|}{0.41} & 54.09    \\ 
MPIN     & \multicolumn{1}{c|}{\best{0.39}} & \multicolumn{1}{c|}{\best{44.92}}    & \multicolumn{1}{c|}{\best{0.38}} & \multicolumn{1}{c|}{\best{44.17}}    & \multicolumn{1}{c|}{\best{0.39}} & \best{44.54}    & \multicolumn{1}{c|}{\best{0.20}} & \multicolumn{1}{c|}{\best{35.89}}    & \multicolumn{1}{c|}{\best{0.21}} & \multicolumn{1}{c|}{\best{36.06}}    & \multicolumn{1}{c|}{\best{0.20}} & \secBest{34.82}    & \multicolumn{1}{c|}{\best{0.24}} & \multicolumn{1}{c|}{\best{10.06}}    & \multicolumn{1}{c|}{\best{0.20}} & \multicolumn{1}{c|}{\best{8.31}}     & \multicolumn{1}{c|}{\secBest{0.16}} & {\secBest{6.67}}     \\ \bottomrule
\end{tabular}
}
\vspace*{-5pt}
\end{table*}

\vspace*{-5pt}
\subsubsection{Overall Efficiency Comparison}


Table~\ref{tab:overall_time} compares the efficiency of all imputation methods in terms of time cost. 
Since the time cost remains consistent across varying missing ratios, we provide a single result for each imputer on the respective dataset.
Notably, methods such as {MEAN}, {FP}, and {MPIN} exhibit considerably lower time costs compared to others. On the other hand, the time-series models {BRITS} and {SAITS} demonstrate the highest time consumption due to their sequential neural network architecture, which includes computationally intensive components like recurrent units or self-attention units during training.
While \MPIN{} is also a neural network model, it capitalizes on parallel processing with its \textsc{MsgProp} layer for graph-structured data, enabling efficient utilization of computational resources and minimizing training time (i.e., the imputation process).

\begin{table}[!ht]
\small
\caption{Overall time cost comparison (unit: seconds).}\label{tab:overall_time}
\resizebox{0.9\columnwidth}{!}{
\begin{tabular}{@{}c|c@{\hspace{4pt}}c@{\hspace{4pt}}c@{\hspace{4pt}}c@{\hspace{4pt}}c@{\hspace{4pt}}c@{\hspace{4pt}}c@{\hspace{4pt}}c@{}}
\toprule
Method     & MEAN & KNN    & MICE  & MF    & FP   & BRITS   & SAITS   & MPIN \\ \midrule
ICU        & 0.05 & 175.25 & 84.84 & 25.28 & 0.32 & 2143.61 & 1985.28 & 3.19 \\ 
Airquality & 0.01 & 81.31  & 0.29  & 1.33  & 0.09 & 1576.1  & 4201.69 & 1.28 \\ 
Wi-Fi      & 0.05 & 0.11   & 42.39 & 4.21  & 0.2  & 7.34    & 9.46    & 2.94 \\ \bottomrule
\end{tabular}
}
\vspace*{-5pt}
\end{table}

Table~\ref{tab:overall_memory} provides a comparison of memory cost among the neural network models, namely {MPIN}, {BRITS}, and {SAITS}. Non-neural network models are excluded in this analysis as they generally exhibit lower effectiveness (as shown in Table~\ref{tab:overall_effective}). The results in Table~\ref{tab:overall_memory} demonstrate that {MPIN} requires clearly less memory compared to {BRITS} and {SAITS}. This distinction arises from the fact that {MPIN}'s \textsc{MsgProp} layer uses a simpler structure than the recurrent unit and self-attention unit used by the others.



\begin{table}[!ht]
\small
\caption{Memory cost (unit: MB).}
\resizebox{0.61\columnwidth}{!}{\begin{tabular}{c|@{\hspace{18pt}}c@{\hspace{18pt}}c@{\hspace{18pt}}c}
\toprule
Method     & BRITS         & SAITS         & MPIN          \\ \midrule
ICU        & 0.374 & 5.256 & \best{0.183} \\ 
Airquality & 0.189 & 5.091 & \best{0.056} \\ 
Wi-Fi      & 23.574 & 12.305 & \best{3.246} \\ \bottomrule
\end{tabular}
}
\label{tab:overall_memory}
\vspace*{-5pt}
\end{table}

Subsequently, we exclude the inferior traditional imputers and narrow our focus to comparing {MPIN} with {BRITS}, {SAITS}, and {FP} as they exhibit superior effectiveness or efficiency. Furthermore, as the MAE results demonstrate similar trends to MRE, they are reported in Section~A.1 in Appendix.

\vspace*{-5pt}
\subsubsection{Effect of Time Window Length}\label{sssec:window_length}

We vary the length of the time window to investigate its impact on imputation. We present the results for \ICU{} and \AIR{} datasets. The results for \WIFI{} follow similar patterns and can be found in Section~A.2 in Appendix.

Referring to Figures~\ref{fig:window_vs_mre_higher} (a) and (b), as the time window becomes longer, the imputation errors of {MPIN}, {SAITS} and {BRITS} decrease. In particular, both {SAITS} and {BRITS} improve more rapidly than {MPIN}. As both {SAITS} and {BRITS} are time series data imputers, they can capture deeper time dependencies and benefit more from longer time series. {In contrast, \MPIN{} does not capture temporal dependencies, and the only benefit to \MPIN{} of a longer time window is to provide more training data. However, this benefit may bring about only slight performance gains, as the training data may already be sufficient.} Still, {MPIN} is the most effective in all cases. 
\begin{figure}[!ht]
\centering
\begin{minipage}[t]{0.19\textwidth}
\centering
\includegraphics[width=\textwidth]{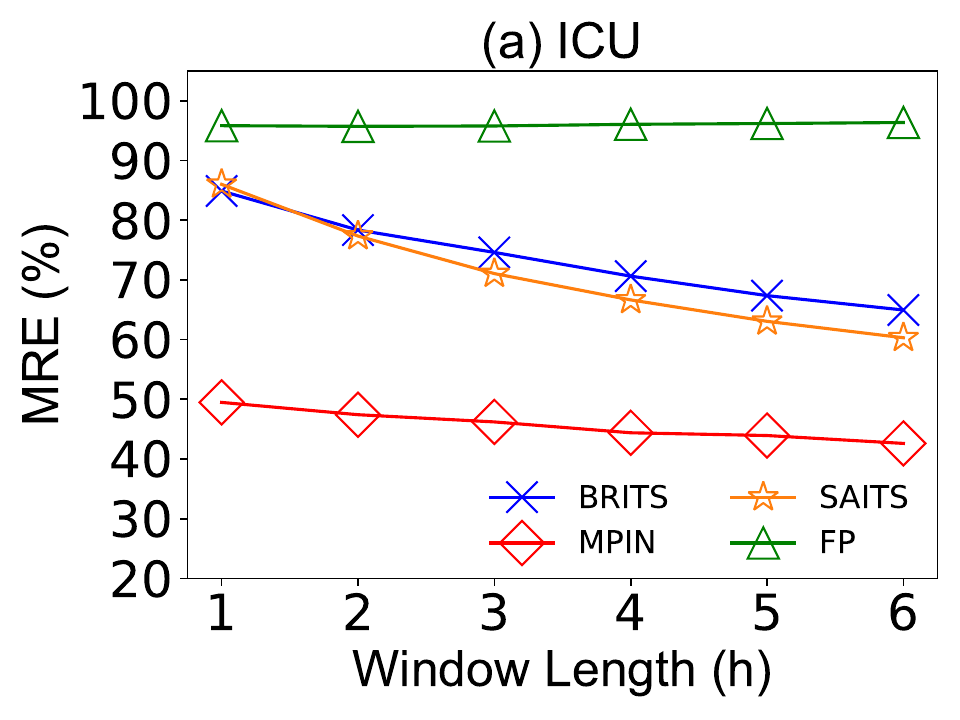}
\end{minipage}
\begin{minipage}[t]{0.19\textwidth}
\centering
\includegraphics[width=\textwidth]{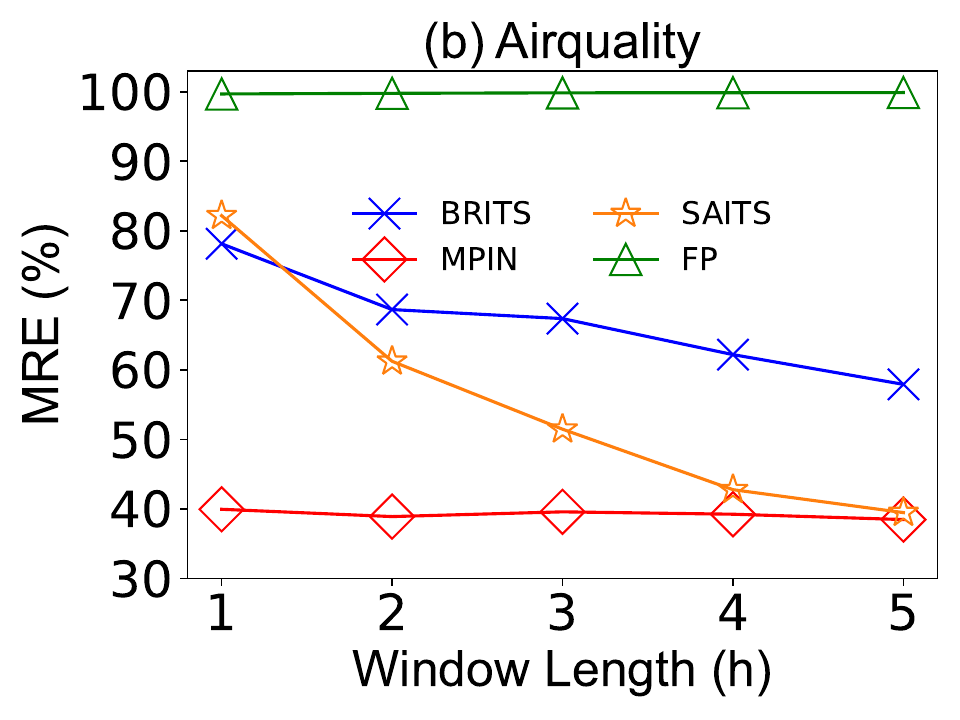}
\end{minipage}
\begin{minipage}[t]{0.19\textwidth}
\centering
\includegraphics[width=\textwidth]{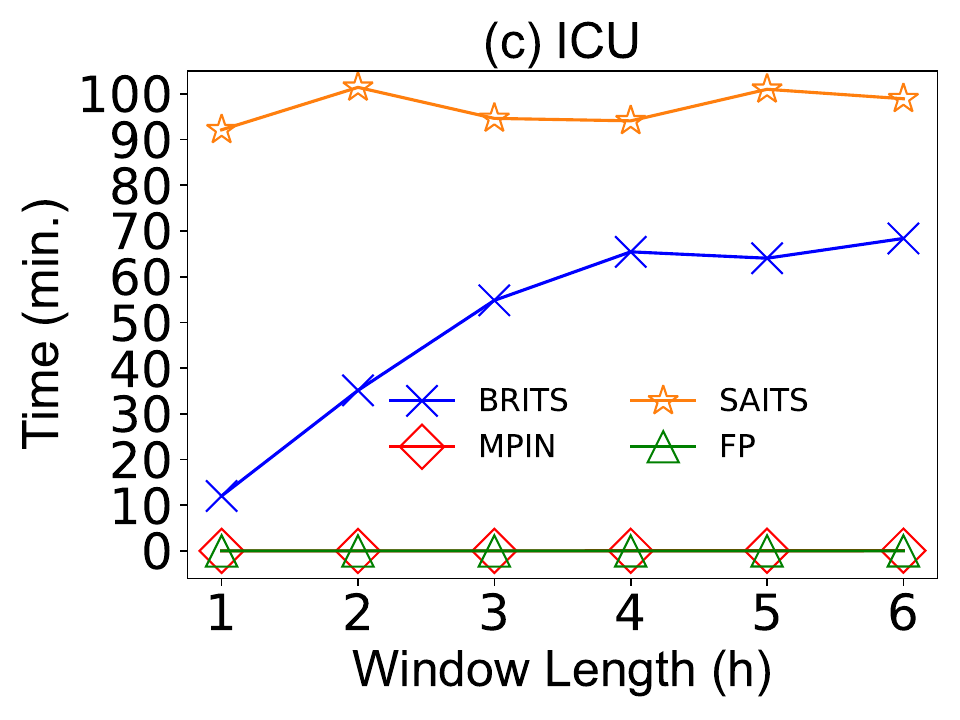}
\end{minipage}
\begin{minipage}[t]{0.19\textwidth}
\centering
\includegraphics[width=\textwidth]{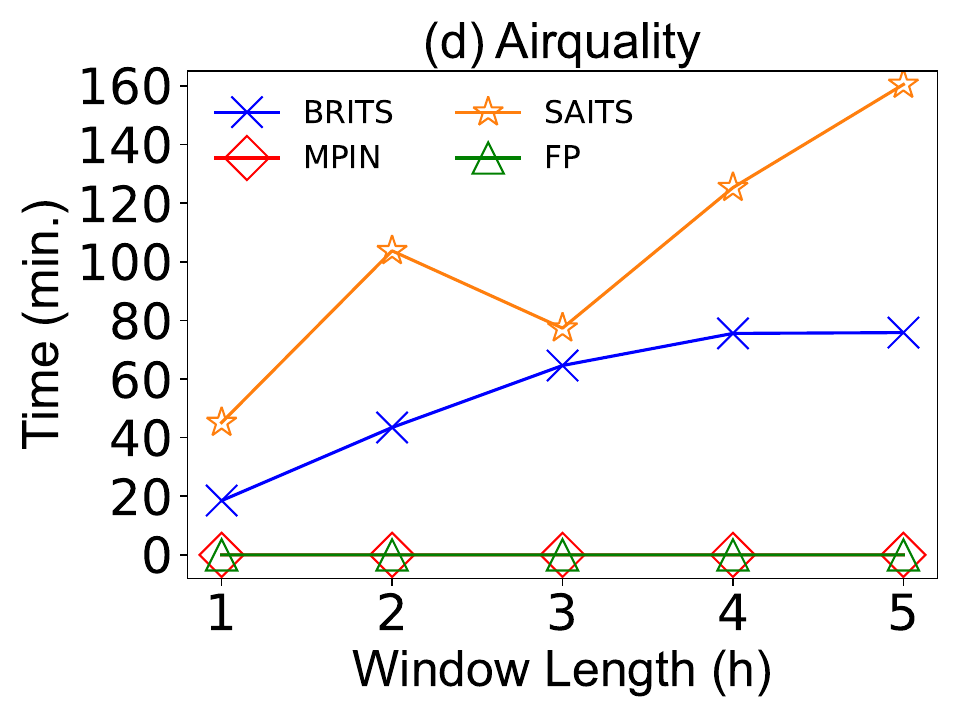}
\end{minipage}
\caption{Effect of window length.}
\label{fig:window_vs_mre_higher}
\end{figure}

Moreover, referring to Figures~\ref{fig:window_vs_mre_higher} (c) and (d), despite fluctuations, the time costs of {SAITS} and {BRITS} increases as the window length increases.
{The longer the time window, the longer the time series that {SAITS} and {BRITS} need to process with their sequential modules.} The fluctuations are due to the random initialization of parameters at the beginning of the training process.

In contrast, {MPIN} maintains consistently low and stable time costs. {This is because \MPIN{} does not process data sequentially but can process graph-structured data in parallel with its \textsc{MsgProp} layer. This enables efficient utilization of computational resources and reduces training time.}
Overall, the time cost of {SAITS} and {BRITS} is by orders of magnitude higher, making them less desirable for online imputation scenarios.
{Also, {FP} exhibits similar efficiency to {MPIN} across different window lengths. However, {FP} lags behind in terms of effectiveness due to its reliance on the simple matrix multiplications in \textsc{FeaProp} that struggle to effectively capture correlations in heterogeneous datasets.}

\vspace*{-5pt}
\subsubsection{Effect of Number of Concurrent Streams}

\begin{figure}[!ht]
\centering
\begin{minipage}[t]{0.21\textwidth}
\centering
\includegraphics[width=0.9\textwidth]{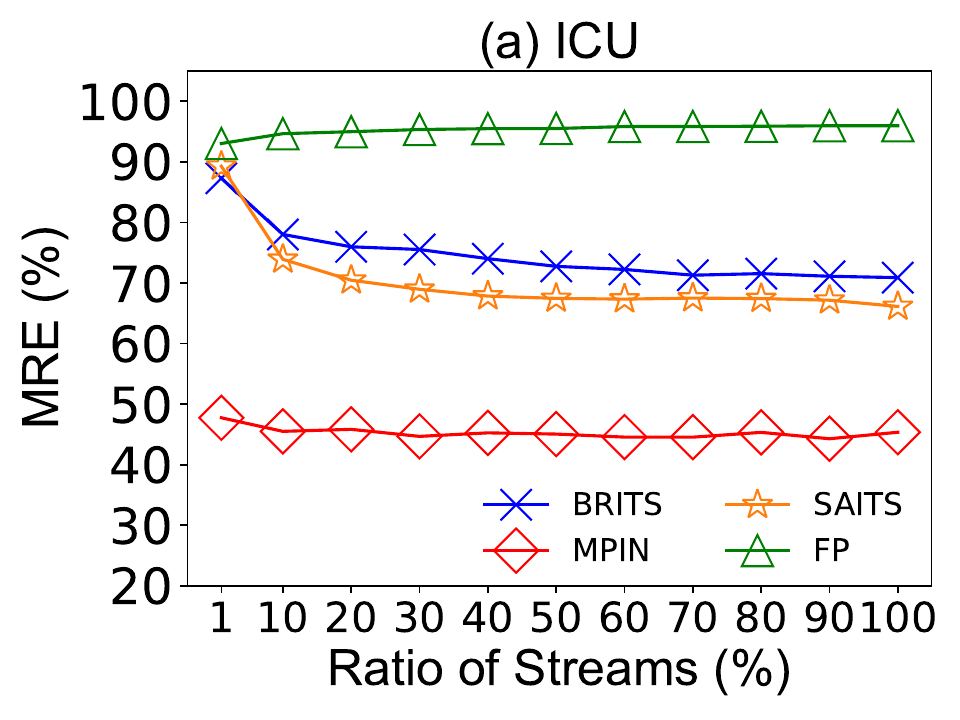}
\end{minipage}
\begin{minipage}[t]{0.21\textwidth}
\centering
\includegraphics[width=0.9\textwidth]{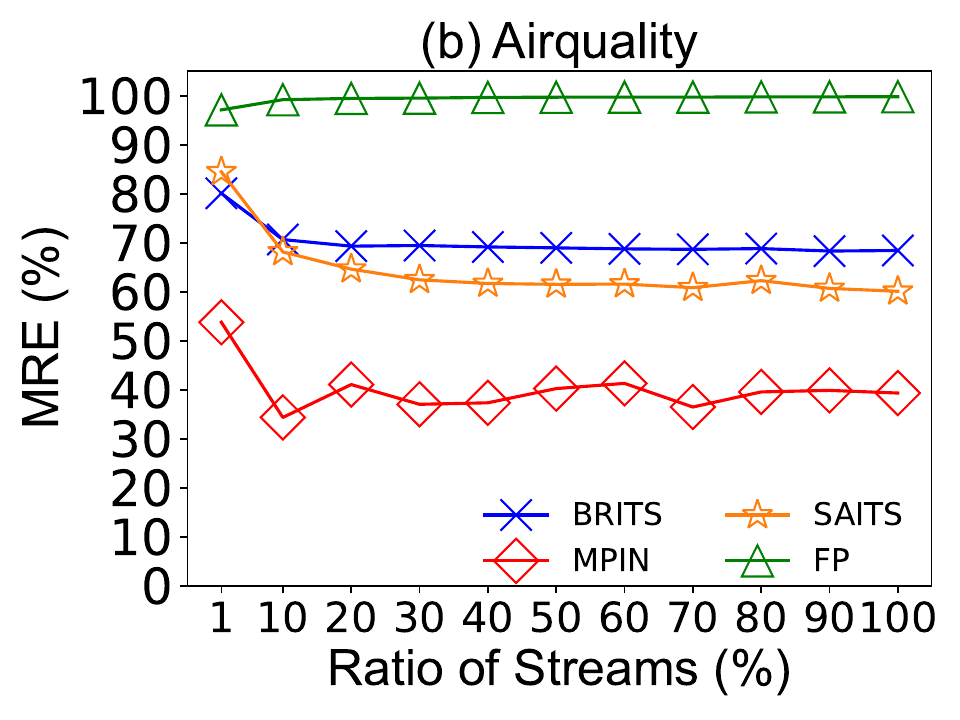}
\end{minipage}
\begin{minipage}[t]{0.21\textwidth}
\centering
\includegraphics[width=0.9\textwidth]{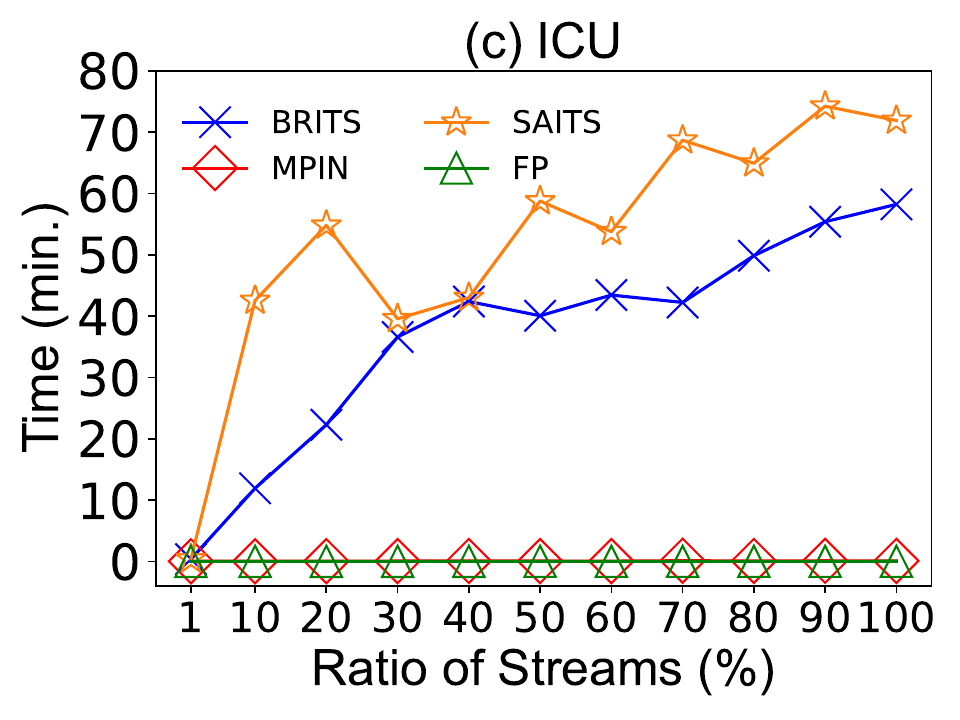}
\end{minipage}
\begin{minipage}[t]{0.21\textwidth}
\centering
\includegraphics[width=0.9\textwidth]{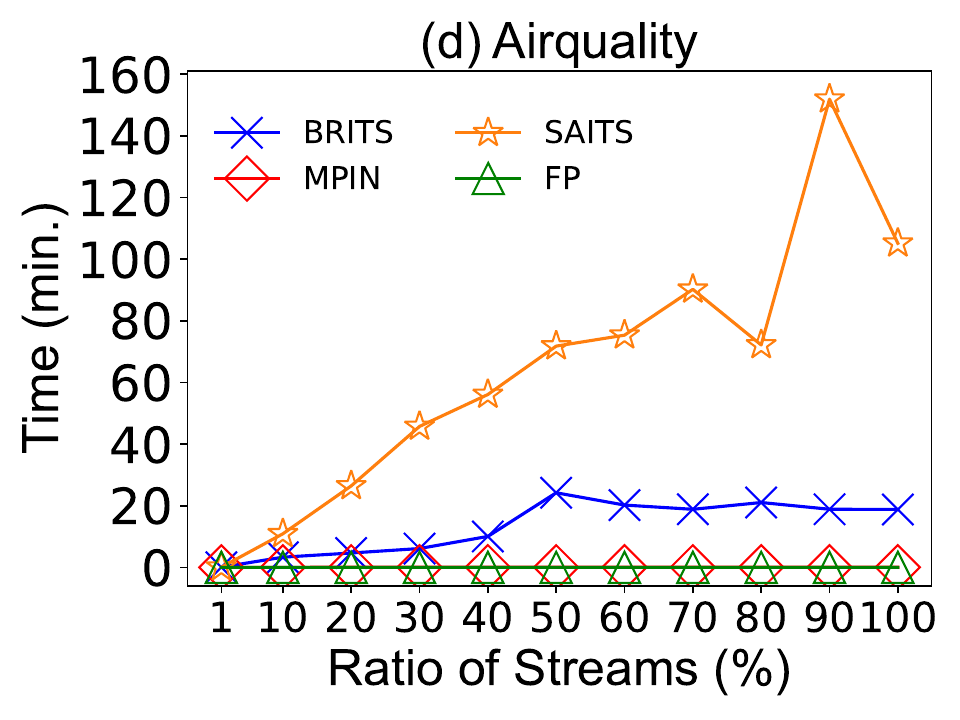}
\end{minipage}
\caption{Effect of concurrent streams. }\label{fig:stream_vs_mre}
\vspace*{-5pt}
\end{figure}

We vary the ratio (number) of concurrent streams to study its impact on imputation quality. 
Referring to Figures~\ref{fig:stream_vs_mre} (a) and (b), as the ratio of streams increases, the MRE of {MPIN}, BRITS, and SAITS initially decreases rapidly and then stabilizes. 
{When the ratio of streams is low, the three neural network models struggle to impute the streams effectively due to the limited amount of data instances for training (imputation).
When the ratio of the streams is higher, their imputation performance is improved due to the additional training data. Furthermore, when the ratio of the streams is excessively high, the advantages diminish as the current data may already be sufficient for training. 
During the whole course, SAITS and BRITS improve more significantly than MPIN. The former are sequential neural network models, and the higher ratio of the concurrent streams means that more sequences can be used by their sequential modules to capture temporal dependencies. 
However, {MPIN} always outperforms BRITS and SAITS even when the ratio of the streams reaches 100\%.   
On the other hand, {FP} shows minimal sensitivity to the ratio of the streams as it is not a neural network model. However, {FP} is always the least effective as it simply relies on adjacency matrix multiplications.}
These findings indicate that the scarcity of data within a time window adversely affects the imputation effectiveness.


Regarding time costs, Figures~\ref{fig:stream_vs_mre} (c) and (d) show that BRITS and SAITS experience degradation as the ratio of concurrent streams increases. Conversely, {MPIN} and FP maintain fast and stable performance.
{The increase in concurrent streams implies that more time series need to be processed by the sequential models BRITS and SAITS, resulting in higher time costs.
In contrast, {MPIN} is non-sequential and can process graph-structured data in parallel.}
In cases where a large data chunk exceeds the available memory, we can split it into medium-sized data chunks and impute each of them individually. This way will minimize the effect of large data chunks and keep the whole process still fast.
%
%

\vspace*{-5pt}
\subsection{Evaluation on Continuous Imputation}
\label{ssec:eva_framework}
%

\noindent\textbf{Methods}. We assess the integration of {MPIN} with the techniques proposed in Section~\ref{sec:incremental} for continuous data stream computation, involving the following methods:
\begin{itemize}[leftmargin=*]
    \item \textbf{\PM{}} periodically calls {MPIN} to impute the missing values of a data chunk at each time window.
    \item \textbf{\DU{}} equips {MPIN} with a data update mechanism to preserve and update valuable data instances for improved imputation at the next window.
    \item \textbf{\MU{}} equips {MPIN} with a model update mechanism that avoids training the model from scratch, and instead resumes training from the best-ever state.
    \item \textbf{\DMU{}} combines both  data update and model update mechanisms with {MPIN}.
\end{itemize}
 We exclude comparisons with other data imputers such as BRITS, as they have been shown to be less effective or efficient (see Section~\ref{ssec:eva_network}). Additionally, the proposed continuous techniques are specifically discussed for {MPIN} only.

\noindent\textbf{Parameter Settings}. Since the proposed continuous techniques primarily address concerns related to data quantity, such as data scarcity within a window, we focus on varying the ratios of concurrent streams to create data chunks of different sizes within the window. The ratios of streams are adjusted at intervals of one order of magnitude, ranging from 0.1\%, 1\%, 10\%, to 100\%. 
The Wi-Fi dataset, which consists of a single stream, is excluded from the variation. Other parameters, such as window length, are kept as defaults.
The reported results are averaged across time windows.  





\subsubsection{Effect of Number of Concurrent Streams}

\begin{figure}[!ht]
\centering
\begin{minipage}[t]{0.21\textwidth}
\centering
\includegraphics[width=\textwidth]{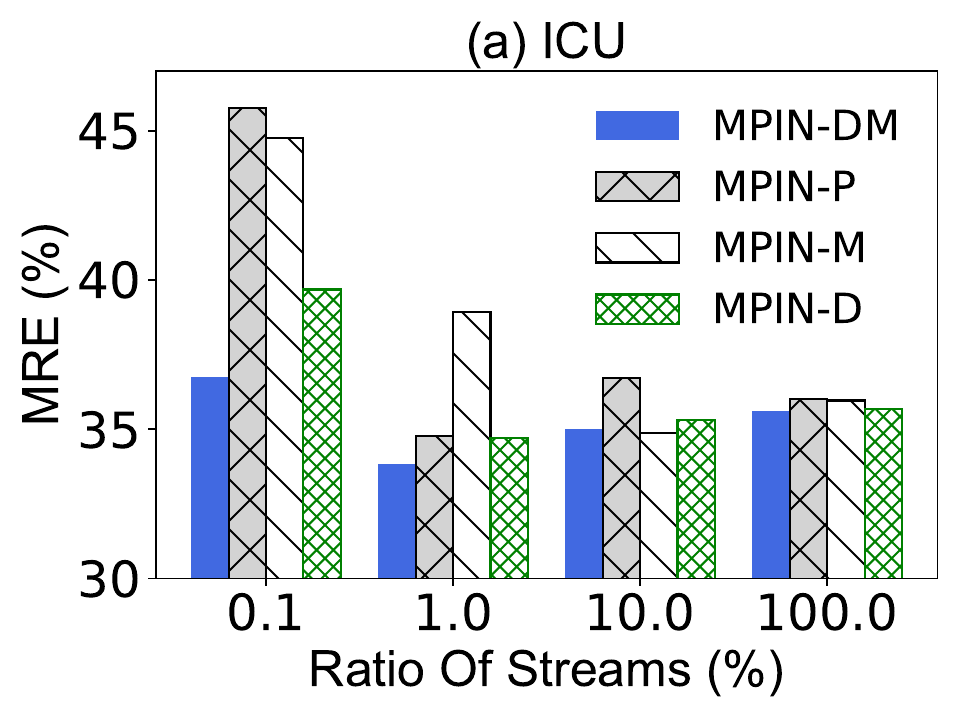}
\end{minipage}
\begin{minipage}[t]{0.21\textwidth}
\centering
\includegraphics[width=\textwidth]{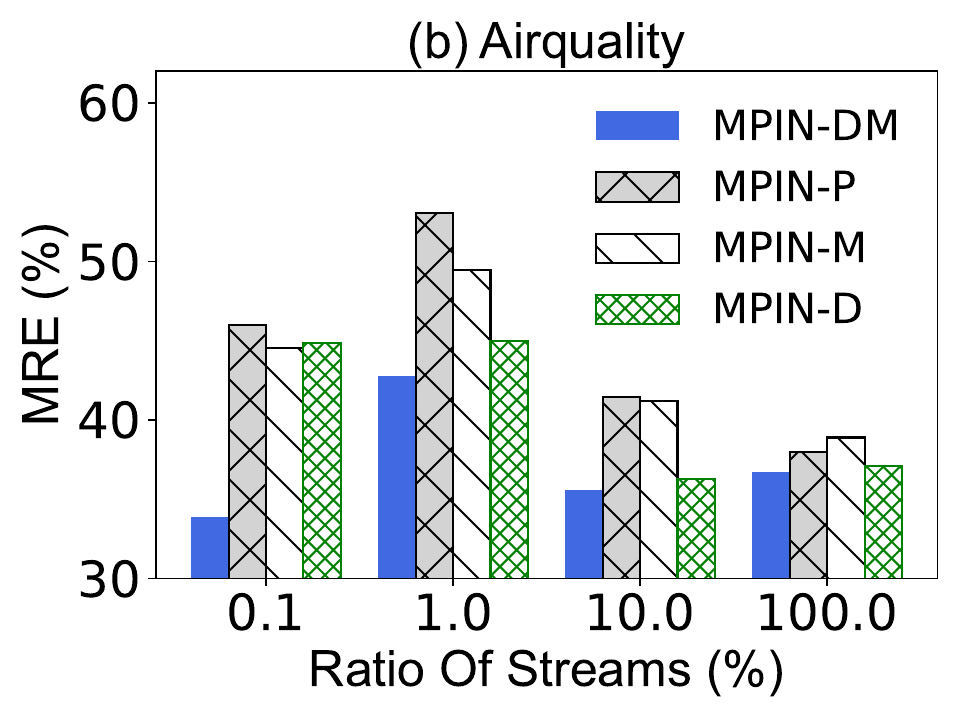}
\end{minipage}
\begin{minipage}[t]{0.21\textwidth}
\centering
\includegraphics[width=\textwidth]{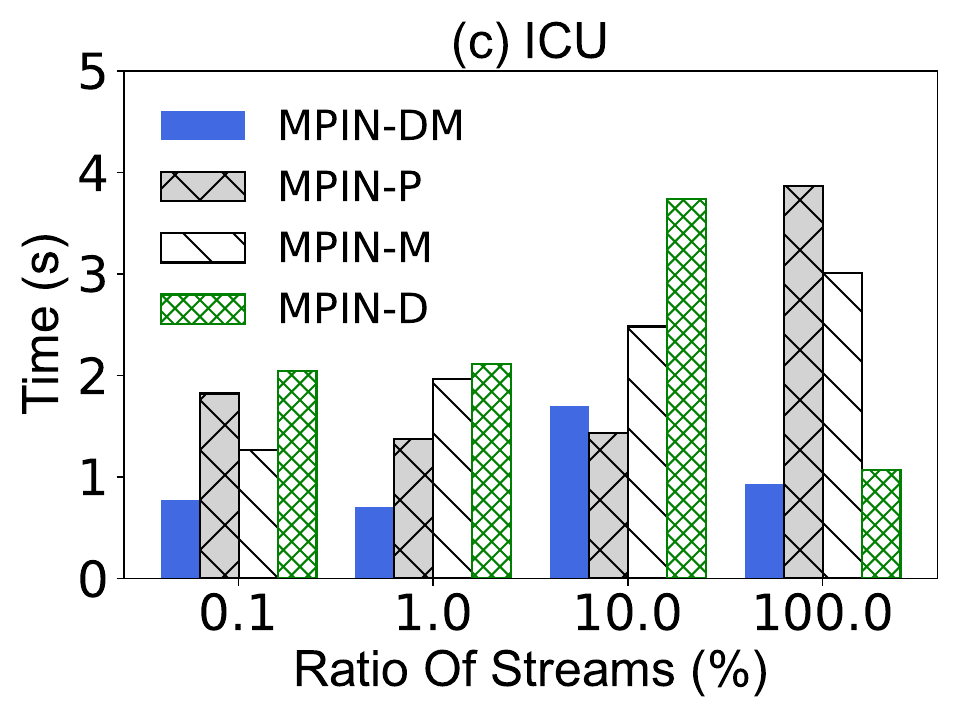}
\end{minipage}
\begin{minipage}[t]{0.21\textwidth}
\centering
\includegraphics[width=\textwidth]{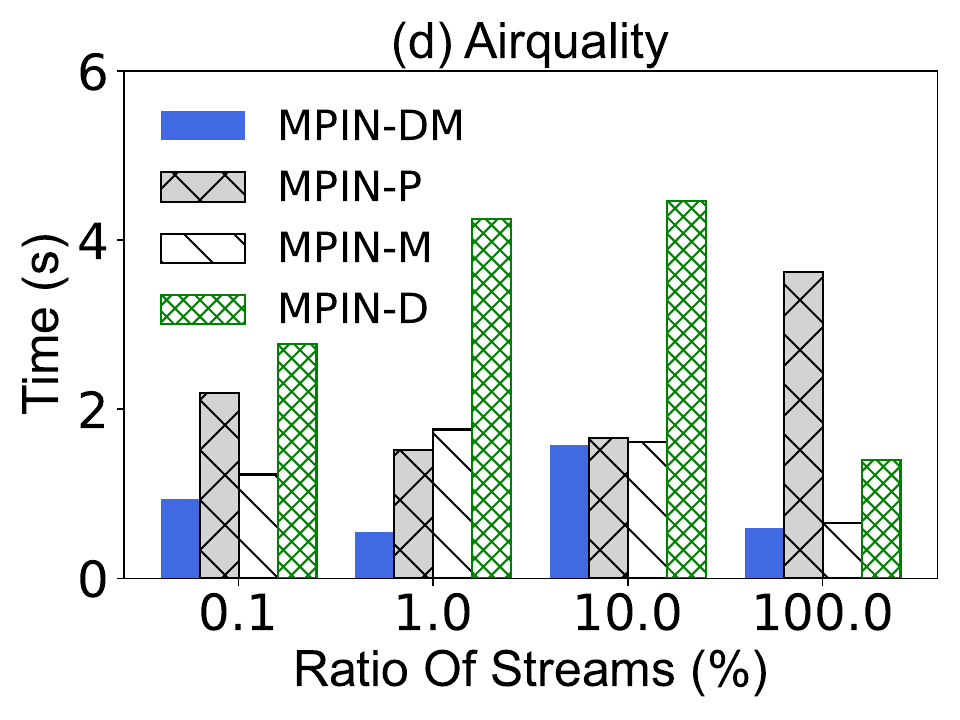}
\end{minipage}
\caption{Continuous imputation on concurrent streams.}\label{fig:effec_of_steam_continuous}
\end{figure}

Referring to Figures~\ref{fig:effec_of_steam_continuous} (a) and (b), both {\DMU{}} and {\DU{}} exhibit lower MRE values compared to {\MU{}} and {\PM{}} in most tested cases. 
This difference becomes more prominent when the ratio of streams is small, such as 0.1\% or 1\%. 
Both {\DMU{}} and {\DU{}} incorporate a data update mechanism, allowing them to utilize valuable past instances and enhance the effectiveness of subsequent imputations. 
When encountering data scarcity within a window, the data update becomes particularly significant as it mitigates the impact of scarcity by augmenting the available data. Moreover, the inclusion of model update in {\DMU{}} further enhances its effectiveness compared to {\DU{}}, as it enables {MPIN} to converge to a better local optimum during training.
{This also explains why {\MU{}} is more effective than {\PM{}} in most cases. However, {\MU{}} may still occasionally be less effective than {\PM{}}, e.g., when ratio=$1.0\%$ on ICU. We attribute this to data distribution drift between windows, which renders the strategy of reusing states from previous windows less effective.} 
Moreover, we notice that the {MRE} of methods does not strictly follow a monotonic trend as the ratio of streams increases or decreases. Nonetheless, in general, the MRE tends to be higher when data is scarce (e.g., 0.1\%) and lower when data is sufficient (e.g., 100\%). Overall, {\DMU{}}, {\DU{}}, and {\MU{}} outperform {\PM{}}, highlighting the effectiveness of the proposed incremental techniques.

On the other hand, referring to Figures~\ref{fig:effec_of_steam_continuous} (c) and (d), in most cases, {\DMU{}} is more efficient than {\DU{}}, and {\MU{}} is more efficient than {\PM{}}. 
This efficiency advantage arises from the model update employed in both {\DMU{}} and {\MU{}}, which allows them to avoid training from scratch and thus reduce time costs. {
 However, in a few cases (e.g., ratio=$1.0\%$), the time cost of MPIN-M exceeds that of MPIN-P. This is likely because the distribution of data between windows may change occasionally, thus making direct reuse of model states from previous windows less effective at training with the data in the current window. Consequently, training MPIN-M (also imputing) converges more slowly than when training from scratch (i.e., MPIN-P). We also notice that MPIN-DM is more efficient than MPIN-M. The data update strategy, i.e., reusing some data from previous windows, can alleviate data distribution drift between windows and can accelerate the model training based on previous model states.}
In general, {\DU{}} is the least efficient due to its training from scratch with additional data (i.e., those valuable data instances). 
{However, when the ratio of the streams is high (i.e., 100\%), the time cost of {\DU{}} is reduced considerably. With large amounts of data, training MPIN may need fewer iterations to converge to a local optimum, which accelerates the training. Moreover, the additional valuable data instances may contribute to this as they provide high-value data instances to speed up the convergence process. 
} 

\subsubsection{Effect of Irregular Sensor Streams}

In previous experiments, the chosen ratio of concurrent streams remained constant across all windows, resulting in data chunks of the same size. 
However, in reality, streams may appear in an irregular or aperiodic manner, leading to sequential data chunks of varying sizes.
%
To simulate this scenario, we introduce variations in the ratio of concurrent streams in \emph{each} time window, including values of 0.1\%, 1\%, 10\%, and 100\%. The imputation results are then averaged across all windows.

\begin{figure}[!ht]
\centering
\begin{minipage}[t]{0.21\textwidth}
\centering
\includegraphics[width=\textwidth]{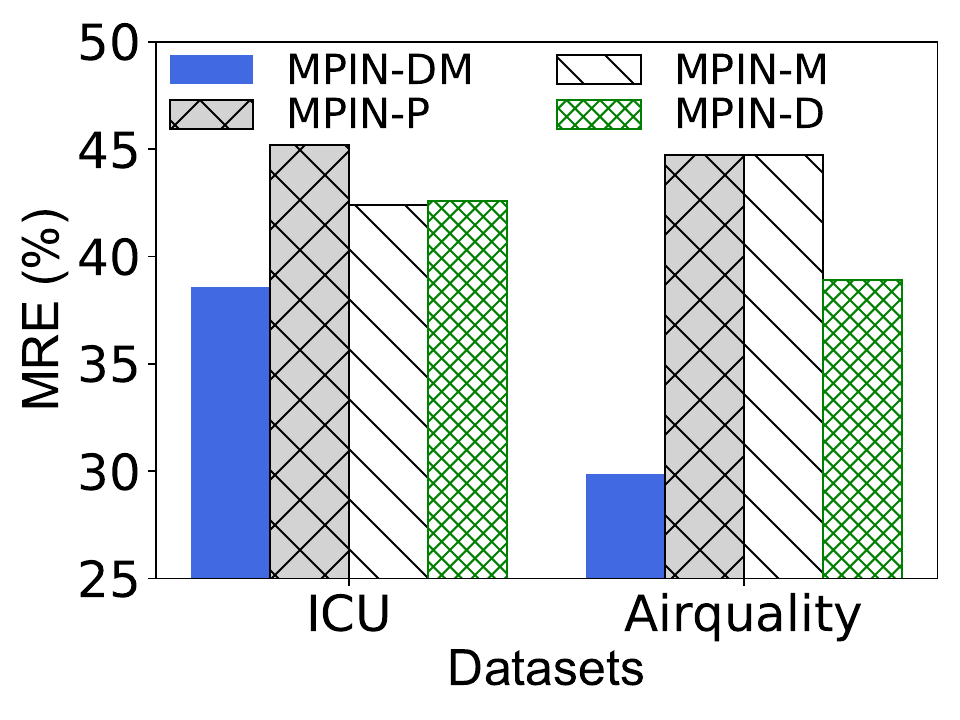}
\end{minipage}
\begin{minipage}[t]{0.21\textwidth}
\centering
\includegraphics[width=\textwidth]{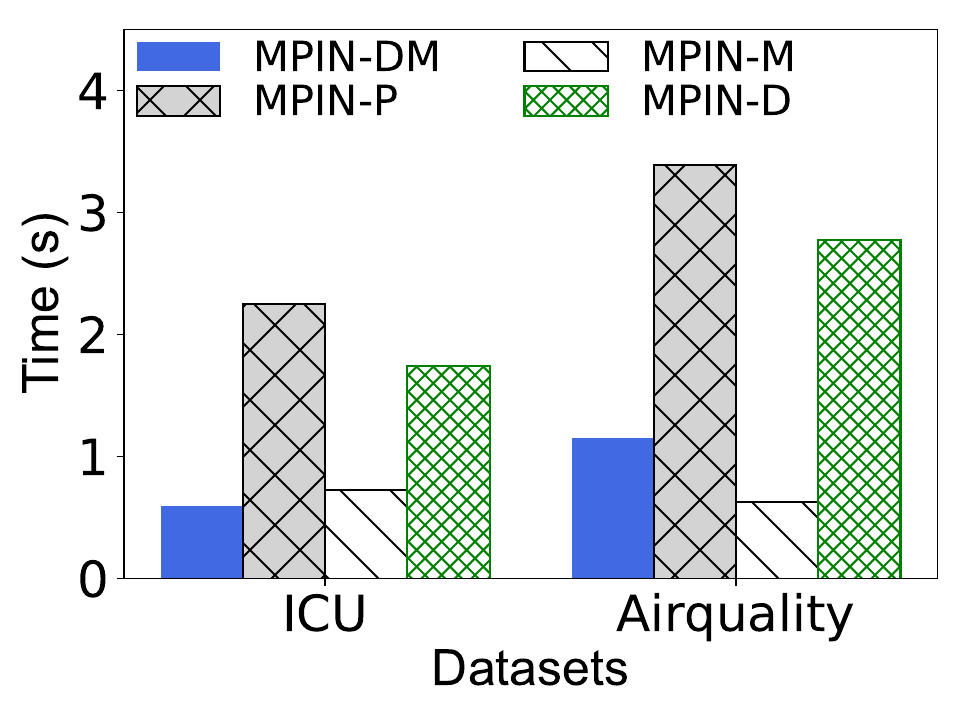}
\end{minipage}
\caption{Performance on irregular streams.}\label{fig:incre_irregular_stream}
\end{figure}

The results in Figure~\ref{fig:incre_irregular_stream} reveal that methods incorporating data update exhibit lower MRE values compared to those without such a mechanism. Specifically, {\DMU{}} outperforms {\MU{}} and {\DU{}} surpasses {\PM{}} in terms of effectiveness.

Regarding time costs, methods using the model update mechanism are more efficient than those without --- {\DMU{}} outperforms {\DU{}} and {\MU{}} outperforms {\PM{}}. Overall, in the presence of irregular streams, the proposed incremental techniques are also helpful. 
Notably, {MPIN} equipped with both data update and model update emerges as the most effective and efficient method.

{Due to the space limit, we present additional experimental results and analyses in the Appendix, including the comparison of  \textsc{MsgProp} with message passing (Section~A.10), ablation study of the combined loss (Section~A.11), and the effect of the transfer mechanism on model update strategy (Section~\ref{sssec:model_update_transfer}).}

\section{Related Work}
\label{sec:related}



\noindent\textbf{Traditional Imputers for Data Streams}.
%
%
Association rules based data imputation methods~\cite{jiang2007estimating,le2005estimating} target wireless sensor networks only and rely heavily on data characteristics for the effectiveness of association rules.
%
Next, KNN-based imputation finds the K nearest neighbors of a data instance in a stream and uses the K neighbors' mean to replace the missing values in the instance ~\cite{zhang2010skif,zhang2012nearest}. However, such simple averaging operations are ineffective on sparse data. 
Furthermore, multiple imputations by chained equation (MICE)~\cite{hastie2009elements} exploits chained equations and is more effective. Still, MICE cannot handle high data sparsity or high dimensionality. 

The top-K case matching method~\cite{wellenzohn2017continuous} finds anchor points that share similar patterns across streams of time series and uses their values to impute missing values. 
A matrix-based imputer~\cite{khayati2020orbits} imputes time series streams by means of an incremental centroid decomposition of the matrix capturing the data. Unlike our work, these two studies consider only homogeneous time series and single-attribute cases. 
An experimental evaluation~\cite{khayati2020mind} compares matrix-based methods~\cite{papadimitriou2005streaming,yu2016temporal,mei2017nonnegative} that use matrix factorization to capture relations among data instances to fill in the missing values. 

Unlike the matrix-based methods, our proposals operate on a similarity graph that benefits data stream imputation because it not only enables capturing correlation but also contains a positive relational bias~\cite{battaglia2018relational,cini2021multivariate}.
An instance in our graph does not need to relate itself to all other instances as in a matrix; instead, it only relates to the most similar neighbors. 
This positive bias provides more accurate and relevant information for missing value imputation. 
%


\noindent\textbf{Neural Network-based Imputers for Data Streams}. 
%
%
Che et al.~\cite{che2018recurrent} use an adapted gated-recurrent unit (GRU) with masking and time-lag mechanisms to impute the missing values in sequences in an iterative fashion. 
Cao et al.~\cite{cao2018brits} propose a  classical bi-directional recurrent neural network for time series data imputation (BRITS). It adopts a transductive learning style and reconstructs missing values by reasoning from observed values in the sequences. 
Studies also exist that combine this RNN-based design with a generative adversarial network (GAN)~\cite{miao2021generative,luo2018multivariate, luo2019e2gan,liu2019naomi} to improve effectiveness. However, this combination may increase the training cost and cause unstable imputation results as GANs are often hard to train~\cite{mescheder2018training}. 
While Cini et al.~\cite{cini2021multivariate} first integrate a message-passing unit with an RNN-based imputer to capture correlations among attributes, they target only homogeneous data.  
Recently, Du et al.~\cite{du2023saits} propose a self-attention-based imputation model to improve the efficiency of imputation. However, the authors only demonstrate that the training time per epoch is reduced, and the full time cost remains high, rendering the model unsuitable for online imputation. 

Overall, the imputation models above suffer from the following drawbacks. First, they are all based on sequential structures (either RNN or attention) and usually need much time for training. 
Second, they only capture intra-sequence temporal dependencies, but not inter-sequence dependencies. 
Third, some of them only work on periodic time series, while streams in practice are often aperiodic. 
In contrast, \MPIN{} operates on a similarity graph and thus can exploit correlations among instances that may come from different streams or different times within a time window. Further, \MPIN{} is amenable to fast training because it enables parallel operations on multiple nodes, thus exploiting available computing resources sufficiently. 

\noindent\textbf{Imputers for Graph-Structured Data}.
%
%
You et al.~\cite{you2020handling} propose to represent tabular data as a bipartite graph, where a node represents an attribute or a label, and then exploit an existing GNN and known labels to impute missing values of feature attributes. 
Spinelli et al.~\cite{spinelli2020missing} propose an encoder-decoder model to impute missing values of graph nodes using label loss. Chen et al.~\cite{chen2022gedi} propose an end-to-end imputation process that takes the known labels and observed feature values into account to impute missing values and predict remaining unknown labels. 
All these imputation methods are designed to exploit (partially) known labels which, however, do not exist in our problem setting. More importantly, these methods apply existing GNN models directly, whereas we enable a message propagation process, with an accompanying theoretical analysis. Finally, the existing methods target only static datasets (e.g., a table) and fall short on data streams. 
{The closest study to ours is feature propagation~\cite{rossi2022unreasonable} (discussed in Section~\ref{ssec:discussion}). That study aims to impute missing values of node features in a pre-existing graph that may lack homophily. In contrast, our approach builds graphs based on the similarity among data instances (see Section~\ref{ssec:similarity_graph}), and the resulting graphs certainly contain homophily to be exploited.}

\section{Conclusion}\label{sec:conclu}

In this work, we have proposed a message propagation imputation network (\MPIN{}) to accurately and efficiently impute missing values in the data instances of a time window in data streams. Moreover, we have proposed a continuous imputation framework with data and model update mechanisms to enable \MPIN{} to support continuous imputation. Extensive experimental studies have demonstrated that the proposed data imputer is generally more accurate and more efficient than competitors and that the continuous framework is effective at continuous imputation on data streams.

For future work, 
it is relevant to apply the proposed techniques to other window types used in stream processing, e.g., sliding windows. Also, 
it is of interest to deploy the proposed techniques in a sensor data streaming system. {Moreover, it is interesting to explore data imputation on heterophilic graphs built on dissimilar data.} 



\clearpage
\balance
\bibliographystyle{ACM-Reference-Format}
\bibliography{cleaning}
\clearpage
\balance
\appendix
\appendixpage

    
\section{Additional Experimental Results}
\label{sec:appendix}

Due to space limit, we include minute details and less significant experimental results and analyses in this appendix.

\subsection{The Results of MAE on ICU and Airquality}\label{ssec:add_exp}

\begin{figure*}
\centering
\vspace*{-3pt}
\begin{minipage}[t]{0.231\textwidth}
\centering
\includegraphics[width=\textwidth]{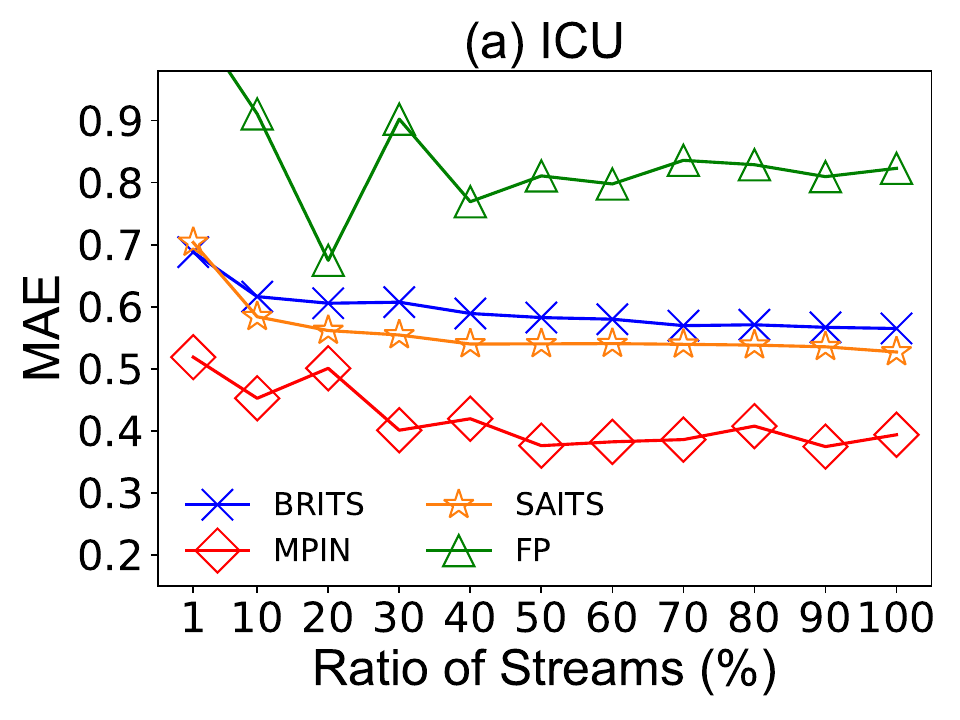}
\end{minipage}
\begin{minipage}[t]{0.231\textwidth}
\centering
\includegraphics[width=\textwidth]{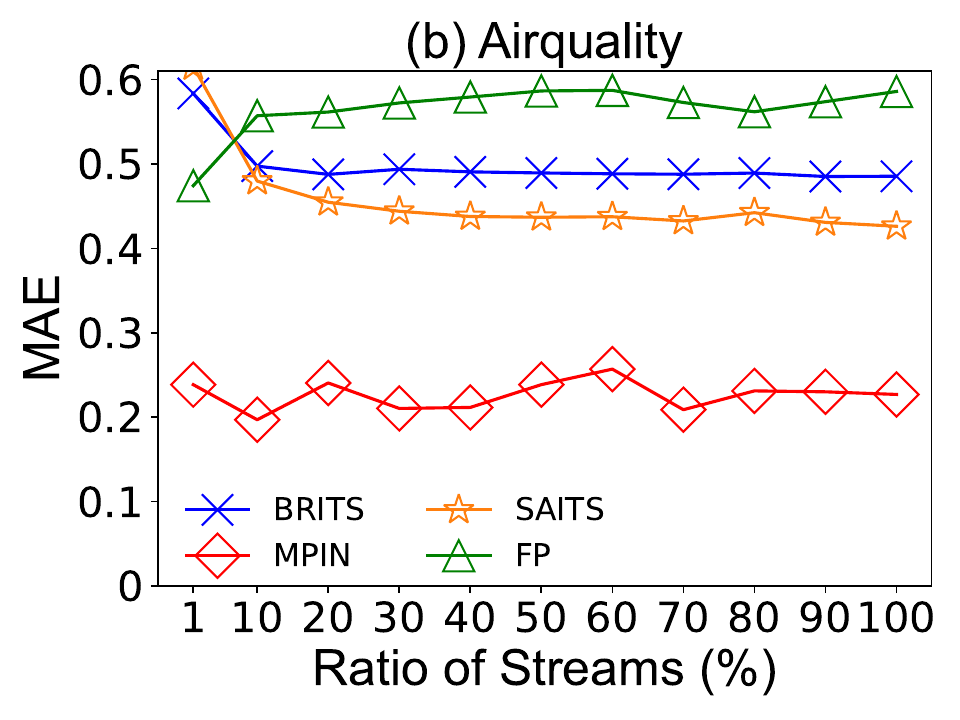}
\end{minipage}
\begin{minipage}[t]{0.231\textwidth}
\centering
\includegraphics[width=\textwidth]{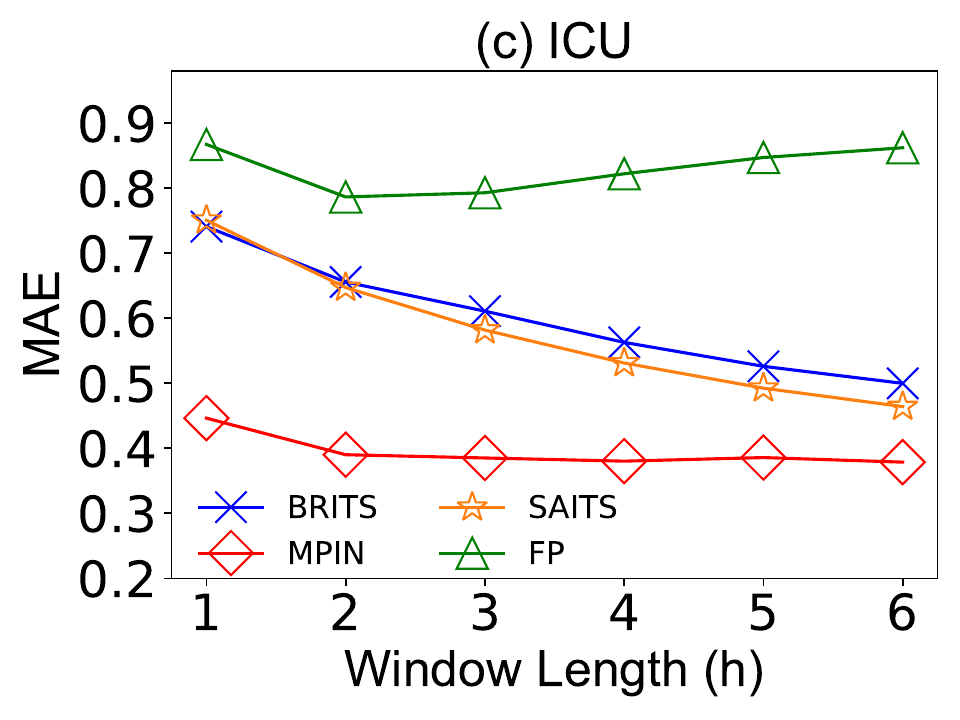}
\end{minipage}
\begin{minipage}[t]{0.231\textwidth}
\centering
\includegraphics[width=\textwidth]{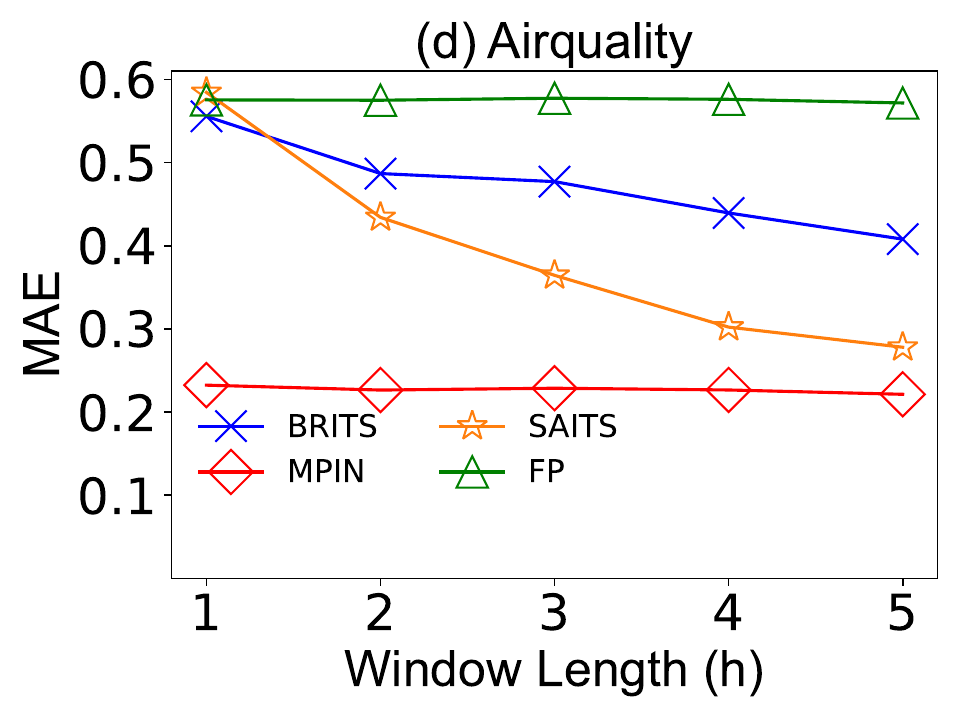}
\end{minipage}
\caption{
MAE results on ICU and Airquality.
}\label{fig:mae_icu_air}
\end{figure*}

We provide MAE results of the ICU and Airquality datasets in Figure~\ref{fig:mae_icu_air}. Overall, the MAE results behave similarly to the MRE results reported in the paper. Referring to Figure~\ref{fig:mae_icu_air} (a) and Figure~\ref{fig:mae_icu_air} (b), MPIN consistently outperforms the other methods across varying ratios of streams. Also, the MAE of the neural network-based methods goes up when the ratio of streams is small (e.g., 1\%). This is due to the lack of training data to impute missing values effectively. In contrast, the effect of the small ratio of streams to FP depends on the dataset. On ICU, the MAE of FP goes up, whereas on Airquality, the MAE goes down. We attribute this to the difference in the sparsity: the sparsity of ICU is much higher than that of Airquality, so FP is more easily affected by the small ratio of concurrent streams on ICU.

Referring to Figure~\ref{fig:mae_icu_air} 
 (c) and Figure~\ref{fig:mae_icu_air} (d), the MAE of all neural network-based methods exhibits a decreasing MAE as the time-window length increases. Usually, a longer window means that more data instances occur in the time window, resulting in more training data for the imputation. Further, the decreasing tendencies of sequential neural networks such as BRITS and SAITS are more pronounced than that of MPIN, since sequential models can capture deeper dependencies from longer sequences. However, a longer time window incurs much more time cost for the sequential models, as reported in the paper. MPIN still achieves the lowest MAE across varying lengths of time windows.

\subsection{Effect of Window Length on Wi-Fi Dataset} 
\begin{figure*}
\centering
\vspace*{-3pt}
\begin{minipage}[t]{0.231\textwidth}
\centering
\includegraphics[width=\textwidth]{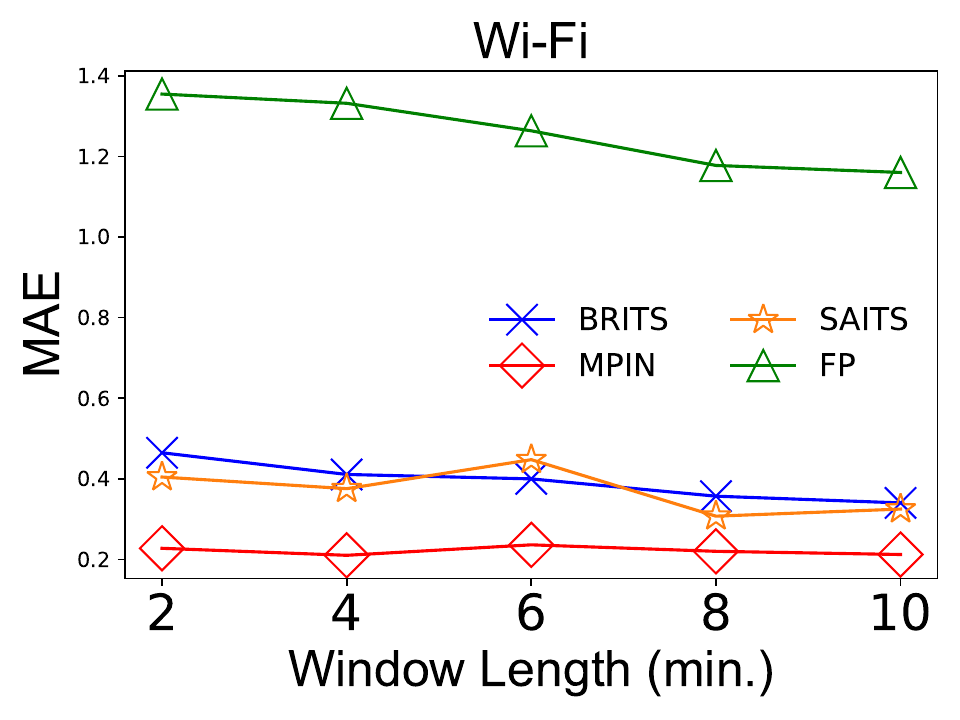}\ExpCaption{MAE vs. window length.}\label{fig:mae_window_wifi}
\end{minipage}
\begin{minipage}[t]{0.231\textwidth}
\centering
\includegraphics[width=\textwidth]{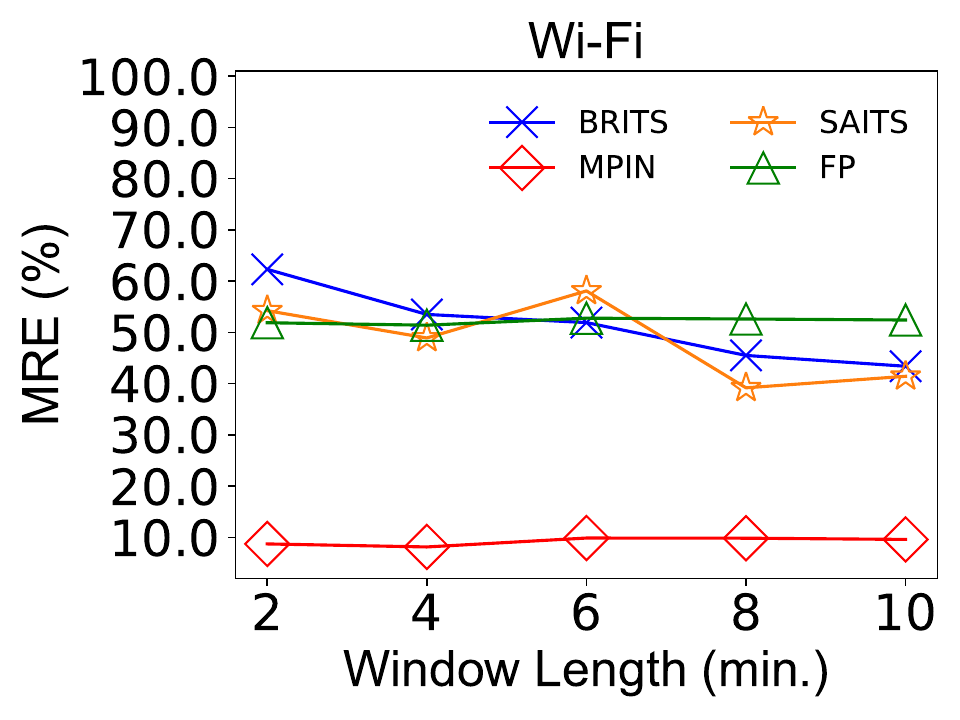}\ExpCaption{MRE vs. window length.}\label{fig:mre_window_wifi}
\end{minipage}
\begin{minipage}[t]{0.231\textwidth}
\centering
\includegraphics[width=\textwidth]{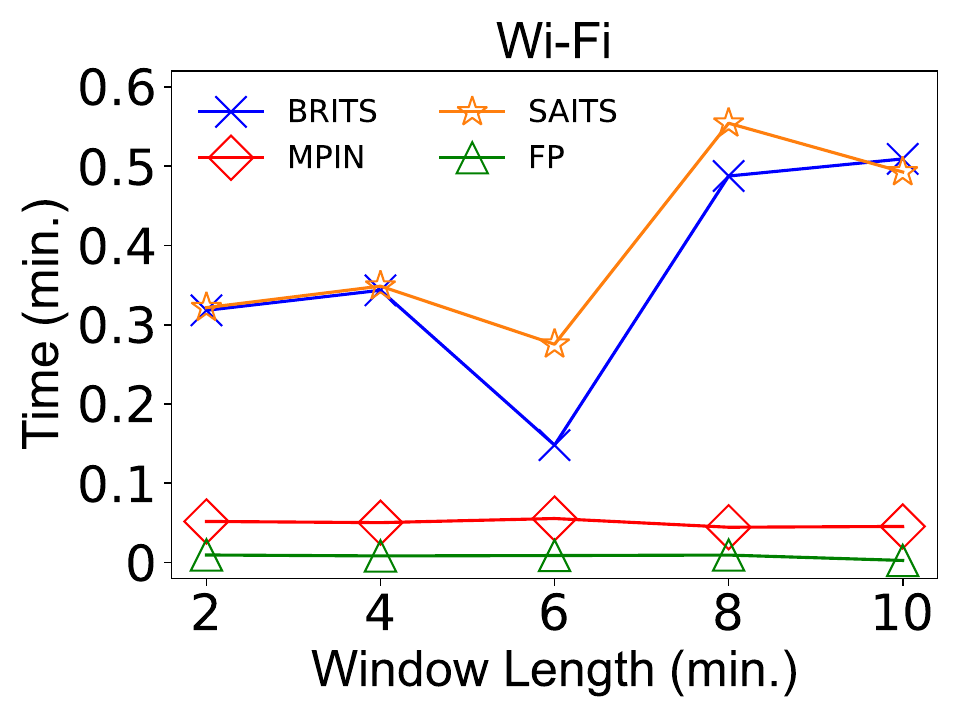}\ExpCaption{Time vs. window length.}\label{fig:time_window_wifi}
\end{minipage}
\begin{minipage}[t]{0.231\textwidth}
\centering
\includegraphics[width=\textwidth]{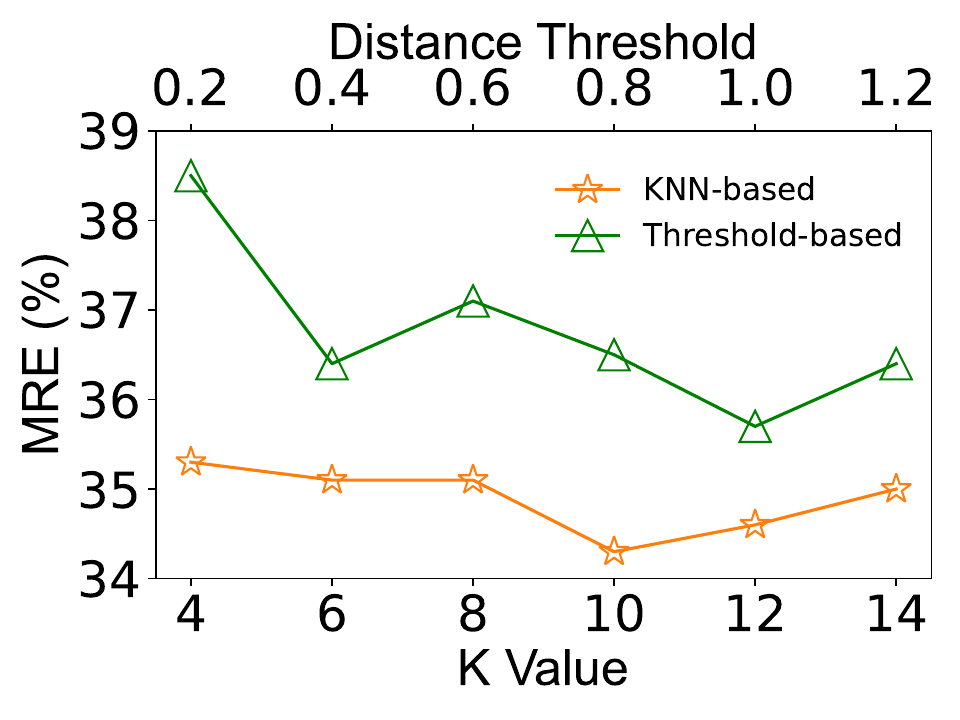}
\ExpCaption{{Effect of methods of graph construction.}}\label{fig:sim_graph_constr}
\end{minipage}
\end{figure*}

As shown in Figures~\ref{fig:mae_window_wifi}, ~\ref{fig:mre_window_wifi}, and ~\ref{fig:time_window_wifi}, we provide experimental results on the Wi-Fi dataset on the effect of window length. Overall, we achieve similar conclusions from Wi-Fi as from the other datasets. First, referring to Figures~\ref{fig:mae_window_wifi} and ~\ref{fig:mre_window_wifi}, MPIN is always the most effective method across varying lengths of time windows. Referring to Figure~\ref{fig:time_window_wifi}, MPIN is much more efficient than the sequential neural network models. Next, FP is the most efficient method since it simply conducts matrix multiplication for imputation. Compared with FP, MPIN takes a little more time but is much better at imputation. Thus, MPIN is the most desirable imputer.


{\subsection{Effect of Methods of Graph Construction}
We compare two methods for building similarity graphs, namely the threshold-based and KNN-based methods using Euclidean distance as the similarity measure. Figure~\ref{fig:sim_graph_constr} shows that the KNN-based method is best in terms of imputation accuracy. This is because it can always ensure that each node (i.e., instance) has K edges, while the threshold-based method can not. Also, with the increase of K, the imputation accuracy first increases and then drops. The best performance is achieved when K=10, so we set K to 10 in the experiments. When K or the distance threshold is low, meaning that there are few edges for each node in the graph, the imputation accuracy is impaired because a node may not be able to capture enough information from neighbors. With the increase of K or the threshold, the imputation accuracy can be improved. 
However, when $K$ or the threshold is too high, the performance degrades. This is because if a node has too many neighboring nodes, some of them may contribute 
 noise because they are not so similar to the target node.}

\subsection{Effect of Similarity Function} Considering the construction of a similarity graph, we use the Euclidean distance to calculate similarity since this yields more effective results compared with using Cosine similarity, as shown in Figure~\ref{fig:ablation_sim}. In all three datasets, a lower MRE is obtained by using Euclidean distance as the similarity function.

\subsection{Effect of Number of \textsc{MsgProp} Layer} 
\begin{figure*}
\centering
\vspace*{-3pt}
\begin{minipage}[t]{0.231\textwidth}
\centering
\includegraphics[width=\textwidth]{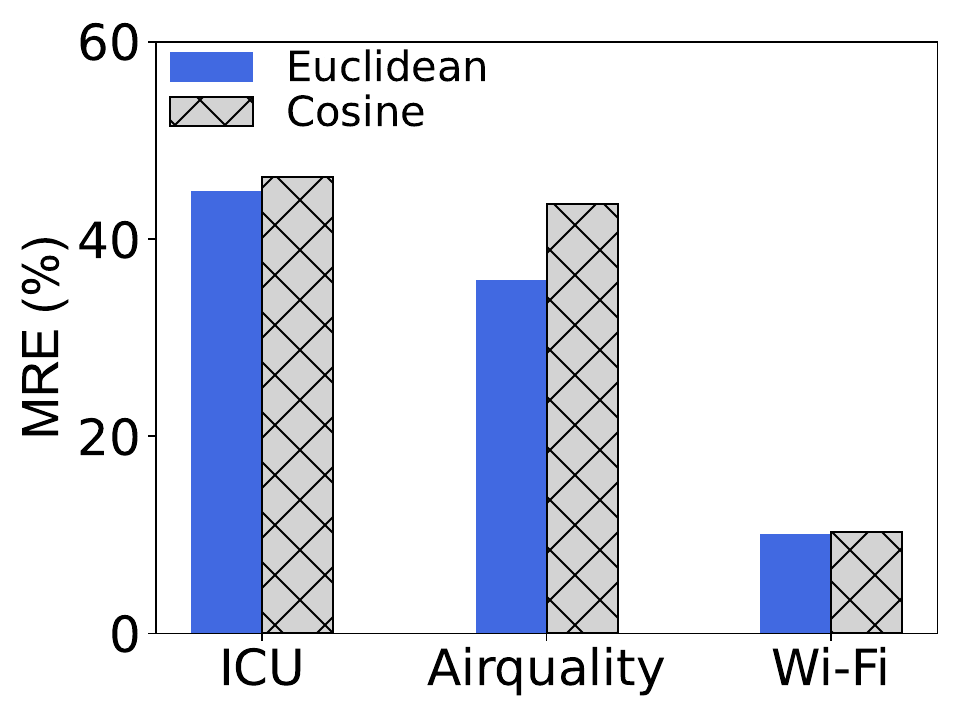}
\ExpCaption{Effect of similarity function.}\label{fig:ablation_sim}
\end{minipage}
\begin{minipage}[t]{0.231\textwidth}
\centering
\includegraphics[width=\textwidth]{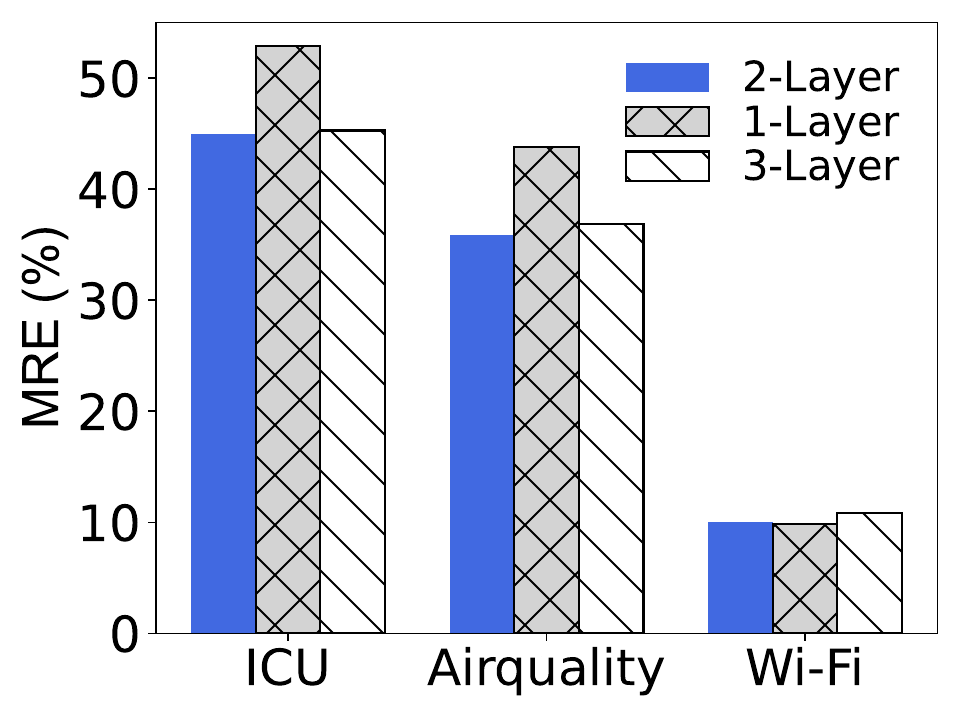}
\ExpCaption{Effect of num of \textsc{MsgProp} layers.}\label{fig:ablation_layer}
\end{minipage}
\begin{minipage}[t]{0.231\textwidth}
\centering
\includegraphics[width=\textwidth]{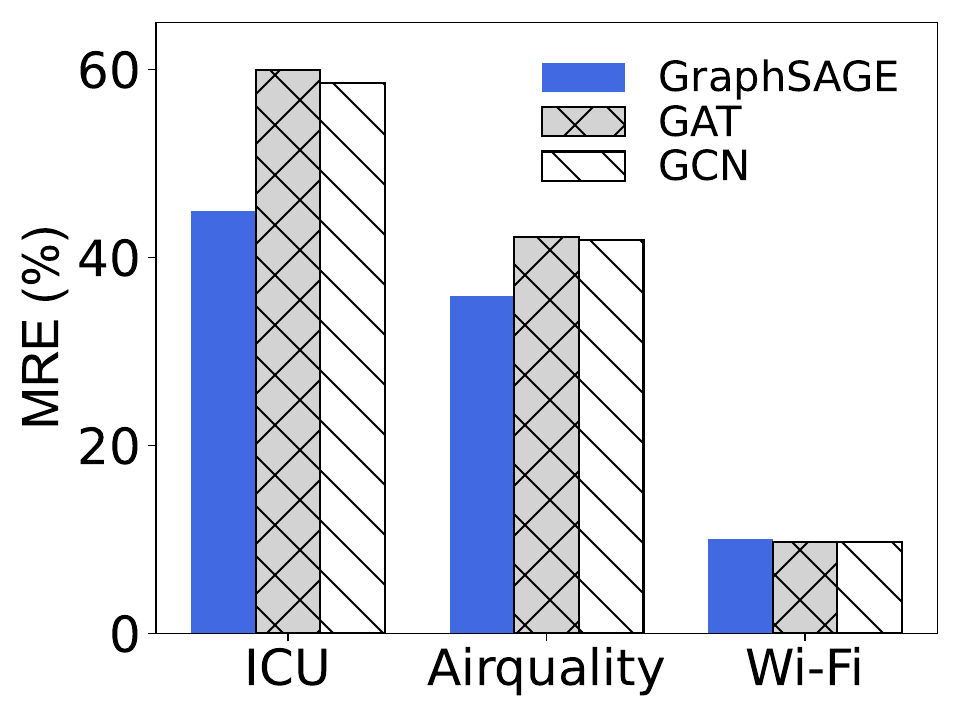}
\ExpCaption{{Effect of types of message passing.}}\label{fig:ablation_basis}
\end{minipage}
\begin{minipage}[t]{0.231\textwidth}
\centering
\includegraphics[width=\textwidth]{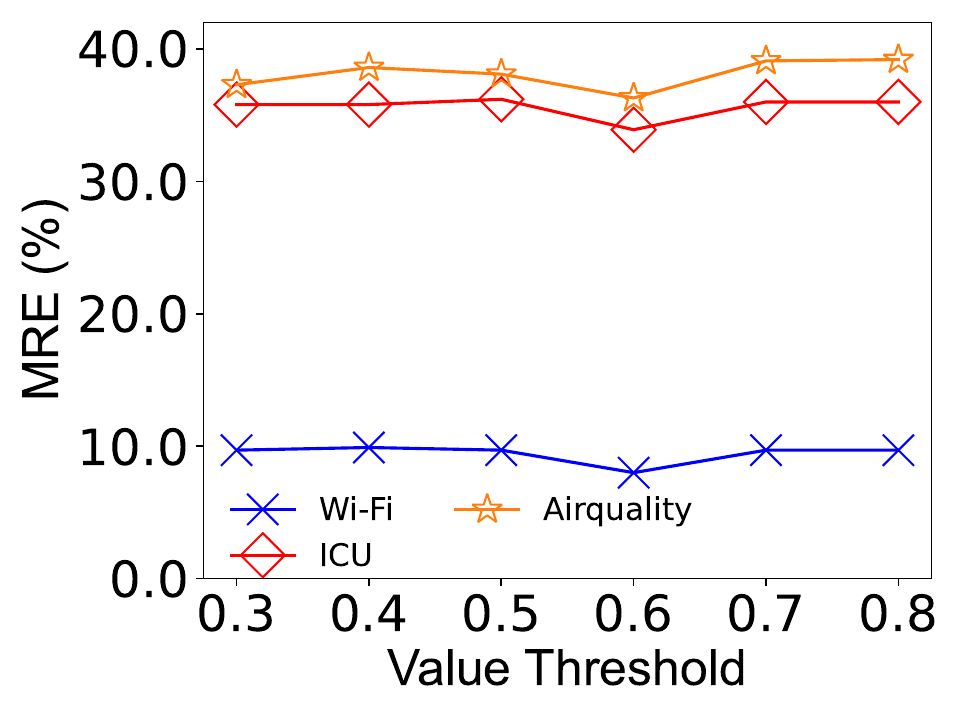}
\ExpCaption{Effect of value threshold.}\label{fig:ablation_thre}
\end{minipage}
\end{figure*}

Referring to Figure~\ref{fig:ablation_layer}, we achieve the best effectiveness on the three datasets when using 2 \textsc{MsgProp} layers. With only one \textsc{MsgProp} 
 layer, MPIN cannot capture the correlations among data instances (i.e., nodes) well and cannot utilize the recursive imputation process. Using 3 or more layers may cause over-smoothing~\cite{chen2020measuring} due to the use of an internal message-passing module. A two-layer MPIN is the best option as it can exploit both the correlations among data instances in the similarity graph and avoid over-smoothing.

{\subsection{Effect of Internal Message Passing Module} We vary the internal message passing module to consider typical message passing modules such as GAT~\cite{velivckovic2017graph}, GCN~\cite{kipf2016semi}, and GraphSAGE~\cite{hamilton2017inductive} to assess their effects on the imputation performance. The results are shown in Figure~\ref{fig:ablation_basis}. Generally, GraphSAGE enables better performance than that of the other two modules. GraphSAGE follows the principle of sampling-and-aggregation that may be better for capturing the correlations among data instances to impute their missing values. }


\subsection{Effect of Value Threshold}\label{ssec:effect_value_thre} To select a proper threshold for the data update strategy, we vary the threshold from 0.3 to 0.8 and report the results of MRE in Figure~\ref{fig:ablation_thre}. As we can see, a lower MRE can be achieved when the threshold is set to around 0.6. Next, the results of MRE appear quite stable across different thresholds, which gives the flexibility to select a proper value threshold.

{\subsection{Performance on Classification Task} We have report on a case study on the ICU dataset. Specifically, we trained a classifier on the imputed sensor dataset to predict whether or not a patient in the ICU would survive~\cite{du2023saits}. We use 3 classification metrics (i.e., AUC, F1-score, and accuracy) to achieve a comprehensive evaluation. The results in Table~\ref{tab:app_classfi}. 
\begin{table}[!htbp]
\small
\centering
\caption{{Performance on Classification Task.}}\label{tab:app_classfi}
\resizebox{0.9\columnwidth}{!}{
\begin{tabular}
{@{}c|c@{\hspace{4pt}}c@{\hspace{4pt}}c@{\hspace{4pt}}c@{\hspace{4pt}}c@{\hspace{4pt}}c@{\hspace{4pt}}c@{\hspace{4pt}}c@{}}
\toprule
Method     & MEAN & KNN    & MICE  & MF    & FP   & BRITS   & SAITS   & MPIN \\ \midrule
AUC       & 0.492 & 0.525 & 0.536 & 0.504 & 0.749 & 0.776 & 0.787 & \best{0.813} \\ 
F1-score & 0.109 & 0.121 & 0.155 & 0.112 & 0.353  & 0.411  & 0.427 & \best{0.459} \\ 
Accuracy      & 0.750  & 0.765 & 0.774 & 0.784 & 0.827 & 0.836 & 0.832 & \best{0.856} \\ \bottomrule
\end{tabular}
}
\end{table}
show that the performance of MPIN in downstream classification tasks is considerably better than those of the other imputation methods. This further demonstrates the significance of MPIN in imputing missing values in sensor data streams. }

{
\subsection{Evaluation on Synthetic Datasets} 
We have conducted additional experiments on two synthetic datasets from an empirical study of imputation~\cite{khayati2020mind}. The performance of each imputation method on the synthetic datasets is shown in Table~\ref{tab:effect_synth}. 
We see that MPIN always achieves much better performance than the other methods across different settings. Only in very few cases is MPIN the second-best imputer. These findings exhibit similar tendencies as those obtained on the real datasets. Thus the study offers further evidence of the effectiveness of MPIN. 
}

\begin{table*}[]
\centering
\caption{{Effectiveness Comparison on Synthetic Datasets (the unit of MRE is \%).}}\label{tab:effect_synth}
\resizebox{0.9\textwidth}{!}{
\begin{tabular}
{@{}c|cccccc|cccccc@{}}
\toprule

Dataset & \multicolumn{6}{c|}{Synthetic Dataset 1}                                                                                                                                                                                                                  & \multicolumn{6}{c}{Synthetic Dataset 2}                                                                                                                                     \\ \cmidrule(r){1-1} \cmidrule(lr){2-7} \cmidrule(lr){8-13} 

Rate  & \multicolumn{2}{c|}{50\%}                                 & \multicolumn{2}{c|}{30\%}                                 & \multicolumn{2}{c|}{10\%}              & \multicolumn{2}{c|}{50\%}                                 & \multicolumn{2}{c|}{30\%}                                 & \multicolumn{2}{c}{10\%}            \\ 
Metrics  & \multicolumn{1}{c|}{MAE}  & \multicolumn{1}{c|}{MRE} & \multicolumn{1}{c|}{MAE}  & \multicolumn{1}{c|}{MRE} & \multicolumn{1}{c|}{MAE}  & MRE  & \multicolumn{1}{c|}{MAE}  & \multicolumn{1}{c|}{MRE} & \multicolumn{1}{c|}{MAE}  & \multicolumn{1}{c|}{MRE} & \multicolumn{1}{c|}{MAE}  & MRE \\ \midrule
MEAN  & \multicolumn{1}{c|}{0.776} & \multicolumn{1}{c|}{99.42} & \multicolumn{1}{c|}{0.773} & \multicolumn{1}{c|}{99.06} & \multicolumn{1}{c|}{0.77}  & 98.63     & \multicolumn{1}{c|}{0.756} & \multicolumn{1}{c|}{99.62} & \multicolumn{1}{c|}{0.754} & \multicolumn{1}{c|}{99.39} & \multicolumn{1}{c|}{0.752} & 99.12     \\
KNN   & \multicolumn{1}{c|}{0.686} & \multicolumn{1}{c|}{87.82} & \multicolumn{1}{c|}{0.625} & \multicolumn{1}{c|}{80.15} & \multicolumn{1}{c|}{0.465} & 59.54     & \multicolumn{1}{c|}{0.609} & \multicolumn{1}{c|}{80.29} & \multicolumn{1}{c|}{0.506} & \multicolumn{1}{c|}{66.69} & \multicolumn{1}{c|}{0.277} & 36.51     \\
MICE  & \multicolumn{1}{c|}{0.657} & \multicolumn{1}{c|}{83.94} & \multicolumn{1}{c|}{0.557} & \multicolumn{1}{c|}{71.21} & \multicolumn{1}{c|}{0.417} & 53.34     & \multicolumn{1}{c|}{0.564} & \multicolumn{1}{c|}{74.36} & \multicolumn{1}{c|}{0.435} & \multicolumn{1}{c|}{57.3}  & \multicolumn{1}{c|}{0.221} & 29.13     \\
MF    & \multicolumn{1}{c|}{0.646} & \multicolumn{1}{c|}{82.73} & \multicolumn{1}{c|}{0.555} & \multicolumn{1}{c|}{71.12} & \multicolumn{1}{c|}{0.418} & 53.61     & \multicolumn{1}{c|}{0.585} & \multicolumn{1}{c|}{77.18} & \multicolumn{1}{c|}{0.477} & \multicolumn{1}{c|}{62.87} & \multicolumn{1}{c|}{0.262} & 34.47     \\
FP    & \multicolumn{1}{c|}{0.591} & \multicolumn{1}{c|}{75.78} & \multicolumn{1}{c|}{\secBest{0.465}} & \multicolumn{1}{c|}{59.55} & \multicolumn{1}{c|}{\secBest{0.395}} & \secBest{50.62}     & \multicolumn{1}{c|}{0.461} & \multicolumn{1}{c|}{60.71} & \multicolumn{1}{c|}{0.31}  & \multicolumn{1}{c|}{40.89} & \multicolumn{1}{c|}{0.213} & 28.04     \\
BRITS & \multicolumn{1}{c|}{\secBest{0.478}} & \multicolumn{1}{c|}{\secBest{58.47}} & \multicolumn{1}{c|}{0.476} & \multicolumn{1}{c|}{\secBest{58.33}} & \multicolumn{1}{c|}{0.469} & 57.48     & \multicolumn{1}{c|}{0.219} & \multicolumn{1}{c|}{26.42} & \multicolumn{1}{c|}{0.199} & \multicolumn{1}{c|}{23.98} & \multicolumn{1}{c|}{\secBest{0.186}} & \secBest{22.41}     \\
SAITS & \multicolumn{1}{c|}{0.518} & \multicolumn{1}{c|}{63.42} & \multicolumn{1}{c|}{0.498} & \multicolumn{1}{c|}{61.02} & \multicolumn{1}{c|}{0.487} & 59.6      & \multicolumn{1}{c|}{\secBest{0.194}} & \multicolumn{1}{c|}{\best{23.4}}  & \multicolumn{1}{c|}{\secBest{0.191}} & \multicolumn{1}{c|}{\secBest{23.06}} & \multicolumn{1}{c|}{0.191} & 23.04     \\
MPIN  & \multicolumn{1}{c|}{\best{0.392}} & \multicolumn{1}{c|}{\best{50.21}} & \multicolumn{1}{c|}{\best{0.382}} & \multicolumn{1}{c|}{\best{48.97}} & \multicolumn{1}{c|}{\best{0.379}} & \best{48.51}     & \multicolumn{1}{c|}{\best{0.185}} & \multicolumn{1}{c|}{\secBest{24.44}} & \multicolumn{1}{c|}{\best{0.156}} & \multicolumn{1}{c|}{\best{20.6}}  & \multicolumn{1}{c|}{\best{0.142}} & \best{18.75}  
\\ \bottomrule
\end{tabular}
}
\end{table*}

\subsection{Message-passing vs. \textsc{MsgProp}}

Though \textsc{MsgProp} uses message passing to capture correlations among instances in the graph, it is distinctive. Essentially, \textsc{MsgProp} equips an additional reconstruction process and computes reconstruction errors for imputation. 
To assess the significance of the distinction empirically, we replace {MPIN}'s \textsc{MsgProp} layer with a message-passing layer so as to discard the intermediate reconstruction process. 
We apply this to the same imputation tasks and show the MRE results in Figure~\ref{fig:ablation_mp}. 
Clearly, the pure message-passing mechanism performs poorly at imputation. In contrast, \textsc{MsgProp} is significantly more effective, as its \textsc{MsgProp} minimizes the Dirichlet energy in the graph, leading to improved reconstructed features (see Section~3.2 in the paper).


\subsection{Ablation Study of Combined Loss}

\begin{figure}[!ht]
\centering
\begin{minipage}[t]{0.22\textwidth}
\centering
\includegraphics[width=0.88\textwidth]{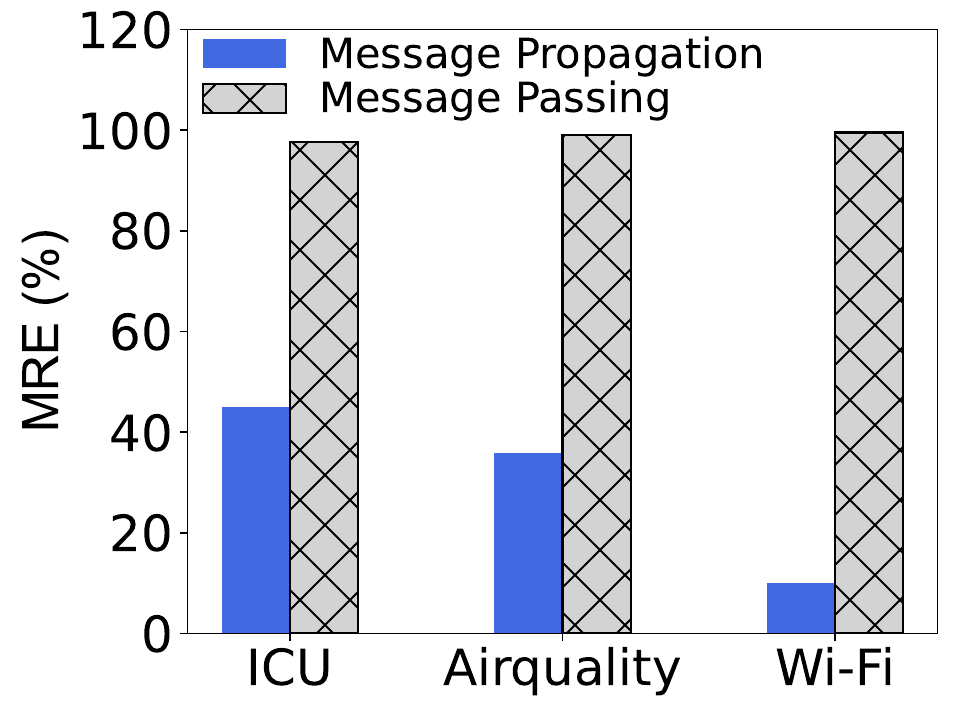}
\ExpCaption{Message mechanism.}\label{fig:ablation_mp}
\end{minipage}
\begin{minipage}[t]{0.22\textwidth}
\centering
\includegraphics[width=0.88\textwidth]{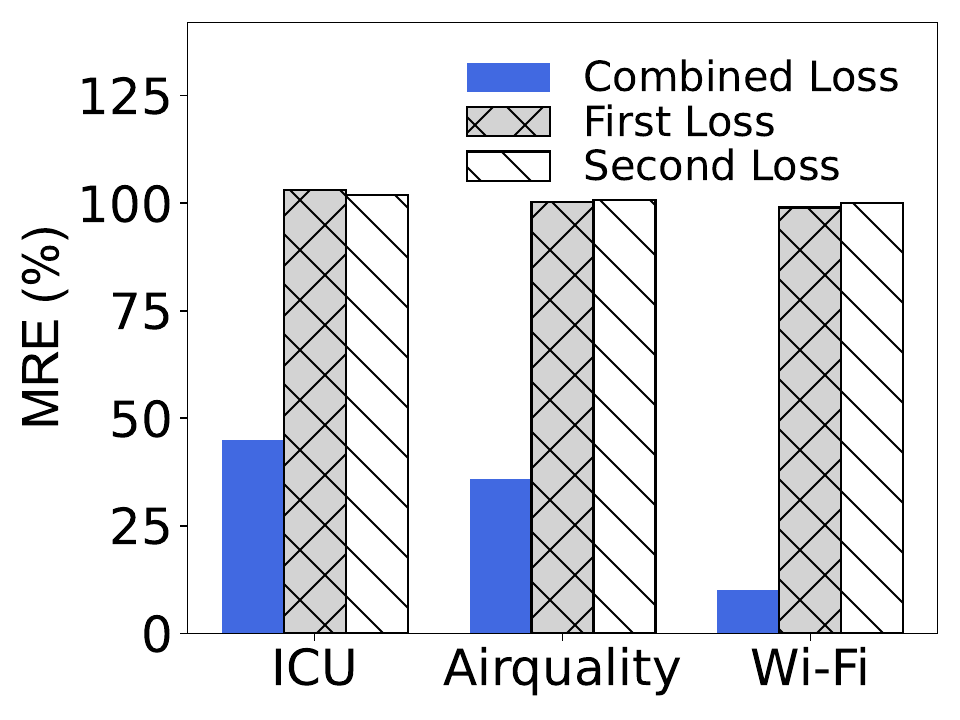}
\ExpCaption{Loss vs. MRE.}\label{fig:ablation_combined_loss}
\end{minipage}
\end{figure}

In {MPIN}, we calculate the loss at the output of each \textsc{MsgProp} layer and use the combined loss for backpropagation. To assess the effectiveness of this design, we compare it with the designs of using only the first-layer loss or using only the second-layer loss. The results reported in Figure~\ref{fig:ablation_combined_loss} show that {MPIN} with its combined loss is much more effective than the two alternatives. 
This is because the reconstruction is conducted on each \textsc{MsgProp} layer and the reconstructed instance of the former layer is taken as the input to the next layer. Thus, it is necessary to consider and handle the error of each layer; otherwise, the model will suffer from accumulated errors.

\begin{figure}[!ht]
\centering
\begin{minipage}[t]{0.21\textwidth}
\centering
\includegraphics[width=\textwidth]{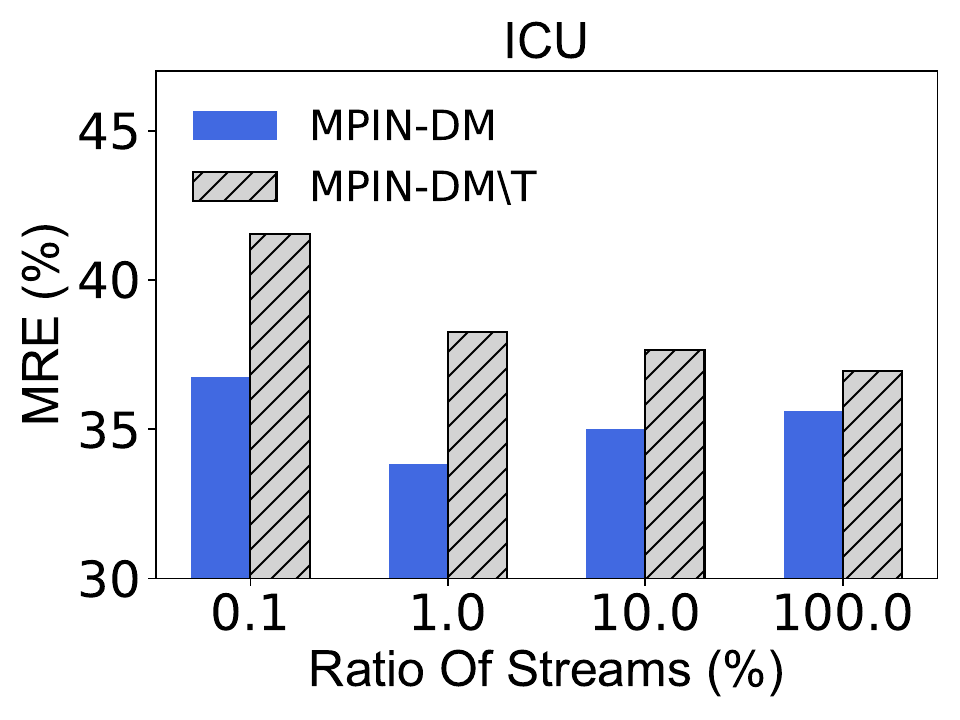}
\end{minipage}
\begin{minipage}[t]{0.21\textwidth}
\centering
\includegraphics[width=\textwidth]{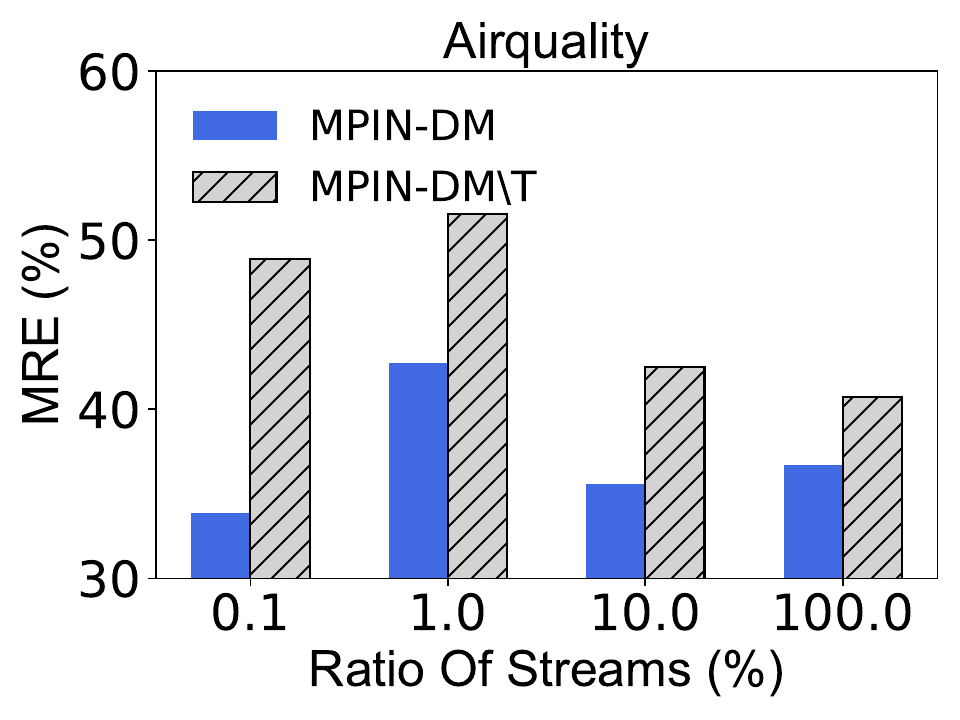}
\end{minipage}
\caption{Effect of transfer mechanism.}\label{fig:mu_transfer}
\end{figure}

\subsection{Effect of Transfer Mechanism of Model Update}\label{sssec:model_update_transfer}

In the design of the model update mechanism (see Figure~6 in the paper), we adopt a selective approach to parameter reuse, rather than blindly reusing all parameters from the best-ever state of {MPIN}. Specifically, we reuse the parameters of the core part, while we retrain the reconstruction part from scratch. This approach draws inspiration from transfer learning.
In our case, the model update process involves transferring the knowledge acquired from one graph to another. The experimental results, shown in Figure~\ref{fig:mu_transfer}, validate the effectiveness of this approach.
In comparison, the {\DMU{}$\backslash$T} method copies all parameters directly from the best-ever state, which is much less effective than our designed model update mechanism. This difference may occur because reusing all parameters can lead to overfitting, while selective parameter reuse based on transfer helps mitigate this issue.

{\subsection{Differences between Label Propagation and MPIN}
The differences between MPIN and label propagation (i.e., LP)~\cite{zhu2002learning} are substantial. MPIN and LP differ not only in how they function but also in the targeted tasks. 
First, most of the differences between FP and MPIN also apply to LP and MPIN, since LP is similar to FP in terms of function. Second, LP and MPIN target different kinds of tasks. As the name implies, label propagation is used for semi-supervised classification tasks on complete data. In contrast, MPIN aims to impute missing attributes in data instances in sensor data streams, which can then be used for classification tasks in downstream applications. Third, LP requires labels to conduct classification tasks, while MPIN requires no labels to perform the process of imputation. }

\section{Proof of Lemmas}\label{sec:lemm_proof}

Here, we provide the proofs of Lemmas 3 and 4.

\subsection{Proof of Lemma 3}\label{ssec:proof_lem3}
\setcounter{lemma}{2}
\begin{lemma}\label{lemma:value_range} 
$\forall \mathbf{x}_i \in V$, we have $0 \leq \varphi(\mathbf{x}_i) \leq \mathtt{D} - 1$.
\end{lemma}

\begin{proof}
Referring to Equation~9 in the paper, the importance score $\varphi(\mathbf{x}_i)$ reaches its maximum value when the first term, i.e., $\mathit{OR}(\mathbf{x}_i)$, reaches its maximum value and the second term, i.e.,  $\frac{1}{(|V|-1)} $ $\sum\nolimits_{\mathbf{x}_k \in V \setminus \mathbf{x}_i}\mathit{OOR}(\mathbf{x}_i, \mathbf{x}_k)$, reaches its minimum value; $\mathbf{x}_i$ reaches its lowest value when the condition is the opposite.

Consider the highest $\varphi(\mathbf{x}_i)$ score. The maximum observation ratio score $\mathit{OR}(\mathbf{x}_i)$ equals $\mathtt{D}$ when $\mathbf{x}_i[d] = 1, \forall 0 \leq d < \mathtt{D}$.
As at least one dimension was observed in any of the other instances $\mathbf{x}_k \in V \setminus \mathbf{x}_i$, the minimum value of the second term cannot be lower than 1.
Therefore, the highest $\varphi(\mathbf{x}_i)$ is achieved as $\mathtt{D} - 1$.

Consider the lowest $\varphi(\mathbf{x}_i)$ score. The minimum observation ratio score $\mathit{OR}(\mathbf{x}_i)$ equals 1 in the case where only one dimension was observed in $\mathbf{x}_i$.
In this case, $\forall \mathbf{x}_k \in V \setminus \mathbf{x}_i$ we have $\mathit{OOR}(\mathbf{x}_i, \mathbf{x}_k) = \mathbf{m}^\mathsf{T}_i \mathbf{m}_k \leq 1$ according to Equation~11.
Therefore, the lowest $\varphi(\mathbf{x}_i)$ must be no less than $1 - 1 = 0$.

To sum up, we have $0 \leq \varphi(\mathbf{x}_i) \leq \mathtt{D} - 1$.
\end{proof} 

\subsection{Proof of Lemma 4}\label{ssec:proof_lem4}

\begin{lemma}\label{lemma:gram_computation}
Given the corresponding gram mask matrix $\mathbf{M}^\text{GM}$, the importance scores  of data instances in the set $V$ can be computed jointly by
\begin{equation}\label{equ:sample_score_matrix}
    \boldsymbol{\varphi} = \frac{(|V| * \operatorname{diag}^{-1}(\mathbf{M}^\text{GM}) -  \mathbf{M}^\text{GM} \cdot \mathbf{1}^{|V| \times 1})}{|V|-1},
\end{equation}
where $\boldsymbol{\varphi} \in \mathbb{R}^{|V| \times 1}$ and $\boldsymbol{\varphi}[i]$ captures the importance score of the $i$-th data instance $\mathbf{x}_i$ in data chunk $\mathcal{X}_a$, $\operatorname{diag}^{-1}(\cdot)$ gets the diagonal vector from the input matrix, and $*$ denotes the element-wise scalar product.
\end{lemma}

\begin{proof}
Following the literature~\cite{drineas2005nystrom}, we have $\mathbf{M}^\text{GM}[i,i] = \mathbf{m}_i^\textsf{T}\mathbf{m}_i$ and $\mathbf{M}^\text{GM}[i,k] = \mathbf{m}_i^\textsf{T}\mathbf{m}_k$.
With $\boldsymbol{\varphi}'$ being the numerator of Equation~\ref{equ:sample_score_matrix}, we have
$$
\begin{aligned}
\boldsymbol{\varphi}' = & \,\, |V| * \operatorname{diag}^{-1}(\mathbf{M}^\text{GM}) -  \mathbf{M}^\text{GM} \cdot \mathbf{1}^{|V| \times 1}  \\
= & [ |V|\mathbf{m}_i^\textsf{T}\mathbf{m}_i - \sum\nolimits_{k=0}^{|V|-1}\mathbf{m}_i^\textsf{T}\mathbf{m}_k : i = 0, \ldots, |V|-1 ] \\
= & [ (|V|-1)\mathbf{m}_i^\textsf{T}\mathbf{m}_i - \sum\nolimits_{k=0 \wedge k \neq i}^{|V|-1}\mathbf{m}_i^\textsf{T}\mathbf{m}_k : i = 0, \ldots, |V|-1 ] \\
= & [ (|V|-1)\mathit{OR}(\mathbf{x}_i) - \sum\nolimits_{\mathbf{x}_k \in V \setminus \mathbf{x}_i} \mathit{OOR}(\mathbf{x}_i, \mathbf{x}_k) : i = 0, \ldots, |V|-1 ].
\end{aligned}
$$
Dividing $\boldsymbol{\varphi}'$ by $(|V|-1)$, we get $\boldsymbol{\varphi} = [\varphi(\mathbf{x}_i): i = 0, \ldots, |V|-1]$.
Therefore, $\boldsymbol{\varphi}[i]$ equals the importance score of $\mathbf{x}_i$ (see Equation~9).
The lemma is thus proved.
\end{proof}

\end{document}